\documentclass[useAMS,usenatbib]{mn2e}

\pdfoutput=1

\usepackage{ifpdf}
\usepackage{amssymb,amsmath}
\usepackage{graphics}
\usepackage{subfigure}
\usepackage[dvips]{graphicx}
\usepackage{epstopdf}
\usepackage{textcomp}
\usepackage{hyperref}
\usepackage[english]{babel}
\usepackage{color}
\usepackage{float}
\usepackage{verbatim}
\usepackage[normalem]{ulem}
\usepackage[T1]{fontenc}  
\usepackage{aecompl}

\title[SN-driven winds in dwarf galaxies]{Supernova-driven winds in simulated dwarf galaxies}

\author[Hu et al.]
{Chia-Yu Hu$^{1}$\thanks{chu@flatironinstitute.org}\\
	$^{1}$Center for Computational Astrophysics, Flatiron Institute, 162 5th Ave NY NY\\
}

\begin{document}
	\maketitle

\begin{abstract}
	
We investigate galactic winds driven by supernova (SN) explosions in an isolated dwarf galaxy using high-resolution (particle mass $m_{\rm gas} = 1{\rm M_\odot}$, number of neighbor $N_{\rm ngb} = 100$) smoothed-particle hydrodynamics simulations that include non-equilibrium cooling and chemistry, individual star formation, stellar feedback and metal enrichment.
Clustered SNe lead to the formation of superbubbles which break out of the disk and vent out hot gas, launching the winds.
We find much weaker winds than what cosmological simulations typically adopt at this mass scale.
At the virial radius, the time-averaged loading factors of mass, momentum and energy are 3, 1 and 0.05, respectively,
and the metal enrichment factor is 1.5.
Winds that escape the halo consist of two populations that differ in their launching temperatures.
Hot gas acquires enough kinetic energy to escape when launched while warm gas does not.
However,
warm gas can be further accelerated by the ram pressure of the subsequently launched hot gas and eventually escape.
The strong interactions between different temperature phases highlight the caveat of extrapolating properties of warm gas to large distances based on its local conditions (e.g. the Bernoulli parameter).
Our convergence study finds that wind properties converge 
when the cooling masses of individual SNe are resolved, which corresponds to $m_{\rm gas}=5 {\rm M_\odot}$ with an injection mass of $500 {\rm M_\odot}$.
The winds weaken dramatically once the SNe become unresolved.
We demonstrate that injecting the terminal momentum of SNe,
a popular sub-grid model in the literature,
fails to capture SN winds irrespective of the inclusion of residual thermal energy.

\end{abstract}

\begin{keywords}
galaxies: dwarf - galaxies: evolution - galaxies: ISM
\vspace{-0.3cm}
\end{keywords}
\

\vspace{-0.5cm}

\section{Introduction}

Galactic winds, galactic scale outflows of gas originated from the interstellar medium (ISM),
play a critical role in the formation and evolution of galaxies (see \citealp{2017ARA&A..55...59N} for a review).
Understanding the origin of galactic winds is still an active field of research.
Broadly speaking, the driving forces fall into two categories:
active galactic nuclei (AGNs) and stellar activities (mostly due to massive stars).
AGN-driven winds are thought to be important for galaxies embedded in massive dark matter halos (halo mass M$_h > 10^{12} {\rm M_\odot}$),
while winds driven by stellar feedback are more important in lower-mass halos (M$_h < 10^{12} {\rm M_\odot}$). 
Among several mechanisms of stellar feedback,
supernova (SN) explosions are of major importance in driving winds due to its dominant energy budget and its bursty nature.

The evolution of a supernova remnant (SNR) can be divided into three phases \citep{1988ApJ...334..252C, 1998ApJ...500..342B, 1998ApJ...500...95T}.
The first phase is the ``free-expansion'' phase,
where the fast-moving stellar ejecta expand unimpededly and gradually sweep up the interstellar medium (ISM).
Once the mass of the swept-up ISM becomes comparable to the ejecta mass,
it enters the energy-conserving Sedov-Taylor phase \citep{1959sdmm.book.....S, 1950RSPSA.201..159T}.
During this phase,
the radial momentum increases as the pressurized bubble expands.
When the thermal energy starts to escape through radiative cooling,
the radial momentum saturates and the SNR enters the momentum-conserving phase.
Recent hydrodynamical simulations have reproduced the analytical results
and it has been shown that the key to faithfully simulate the dynamical impact of a SN on its ambient ISM is to resolve its energy-conserving phase \citep{2015MNRAS.450..504M, 2015ApJ...802...99K, 2015ApJ...809...69S}.

Modeling galactic winds originating from a multi-phase ISM with numerical simulations is challenging because of the high resolution it requires.
This is currently not feasible for large-scale cosmological simulations, and is expected to remain so in near future.
Therefore, phenomenological approaches have to be taken,
making these galactic winds an input of the model rather than a prediction (see \citealp{2015ARA&A..53...51S} for a review).

While simulating AGN-driven winds that resolve the entire dynamic range of relevance seem to remain out of reach,
it has become possible recently to simulate SN-driven winds in a representative volume of a galaxy with the required resolution and relevant ISM physics.
The common approach is to simulate a patch of the galaxy (typically a Milky Way-like galaxy) with a gravitationally stratified box,
assuming periodic boundary conditions along the disk plane and open boundaries in perpendicular directions \citep{2012ApJ...750..104H, 2016ApJ...816L..19G, 2016MNRAS.456.3432G, 2017MNRAS.466.1903G, 2018ApJ...853..173K}.
While significant processes have been made,
the assumption of periodicity in this type of simulations renders them insufficient to predict the large-scale properties of winds (see Appendix C in \citealp{2016MNRAS.459.2311M} for a discussion),
which can only be inferred indirectly \citep{2017ApJ...841..101L, 2018ApJ...853..173K}.

Star-forming dwarf galaxies,
besides being interesting objects on their own,
provide a unique opportunity for galactic scale simulations (i.e., simulating the entire galaxy rather than just a patch of it) with the required resolution thanks to their small sizes \citep{2016MNRAS.458.3528H, 2017MNRAS.471.2151H}.
They serve as an ideal laboratory where numerical experiments can be conducted so as to strengthen our understanding of not only the physics of the ISM but also the numerical model we adopt.
High-resolution galactic scale simulations can therefore provide valuable guidelines for developing sub-grid models for cosmological simulations.

In this work,
we present high-resolution ($1 {\rm M_\odot}$) hydrodynamical simulations of SN-driven winds originated from the multi-phase ISM of an isolated dwarf galaxy from pc-scales where individual SNe are resolved all the way to the virial radius of the halo.
We quantify the amount of mass, momentum, energy and metals carried by the winds and investigate how winds escape the halo,
taking advantage of the Lagrangian nature of our code.
We study the convergence properties of winds by gradually coarsening the resolution, 
and we show that the resolution requirement to obtain converged wind properties in our simulations is $5 {\rm M_\odot}$,
which is the scale that starts to resolve individual SNe.
Finally,
we explore different schemes for SN feedback and its impact on galactic winds.
We demonstrate that the wind properties are insensitive to the SN schemes as long as the SNe are resolved.
In addition, 
we find that injecting the terminal momentum of SNe,
a widely adopted sub-grid model,
is unable to improve the convergence due to its assumption that the thermal energy has been radiated away right after the injection,
and the fact that winds are driven by thermal pressure.

This paper is structured as follows.
In section \ref{sec:method}, we present the adopted numerical methods and the simulation setup.
In section \ref{sec:results}, we discuss the wind properties in our simulations,
present the convergence study and demonstrate the effect of different SN schemes.
In section \ref{sec:discussion}, we discuss our results comparing to other studies in the literature.
In section \ref{sec:summary}, we summarize our work.

\section{Numerical Method}\label{sec:method} 

The numerical methods we adopt in this paper are largely based on \citet{2016MNRAS.458.3528H} and \citet{2017MNRAS.471.2151H}, 
with a number of revisions.
Here we briefly summarize the methods and give a detailed description of the revised aspects.

For gravity and hydrodynamics,
we use {\sc Gadget-3}  \citep{2005MNRAS.364.1105S} with a modified implementation of smoothed particle hydrodynamics (SPH) \citep{2014MNRAS.443.1173H}.
We follow non-equilibrium cooling and heating processes and a chemistry network for molecular hydrogen based on \citet{1997ApJ...482..796N}, \citet{2007ApJS..169..239G} and \citet{2012MNRAS.421..116G}.

We adopt a stochastic star formation recipe:
for each gas particle that is eligible for star formation on a timestep $\Delta t$,
it has a probability of $\epsilon_{\rm sf} \Delta t / t_{\rm ff}$ ($t_{\rm ff}$ is the gas free-fall time and $\epsilon_{\rm sf}$ is an efficiency parameter) to be converted into a star particle,
with its position, velocity, mass and metallicity unchanged but acts like a collisionless particle.
We assume $\epsilon_{\rm sf} = 0.5$ in this work.
This high value of efficiency is justified by our high resolution which is able to resolve the structure of dense clouds.
Star formation occurs only when (\textit{i}) local velocity divergence becomes negative and (\textit{ii}) the local Jeans mass of the gas particle drops below the total mass within an SPH smoothing kernel (the ``kernel mass'').
At our fiducial resolution,
each star particle represents an actual star whose mass is drawn from a Kroupa stellar initial mass function (IMF) \citep{2001MNRAS.322..231K} rather than a stellar population.
We implement a revised sampling method that will be described in details in Section \ref{sec:imf_sample}.

We consider three types of stellar feedback:
photoelectric heating, photoionization and SN explosions.
Photoelectric heating is self-consistently calculated using the FUV fluxes originated from star particles. 
The FUV radiation is assumed to be optically thin between the stellar sources and gas particles,
albeit with a local shielding treatment using the smoothing length of the gas particles.
This is a fair approximation in dwarf galaxies that have low dust-to-gas ratios like ours.
Photoionization is based on a Str\"{o}mgren type approximation with an iterative method that can cope with overlapping HII regions.

SN feedback includes both supernova type II (SNII) and type Ia (SNIa).
All massive stars ($\geqq 8 {\rm M_\odot}$) will explode as SNII at the end of their lifetimes.
A small fraction of low-mass stars ($< 8 {\rm M_\odot}$) will explode as SNIa following a delay-time distribution (see Section \ref{sec:dtd}).
We adopt a revised implementation of SN feedback which will be described in Section \ref{sec:SNfeedback}.

We track eleven individual elements including hydrogen and helium which evolve as passive scalars and contribute to radiative cooling.
We include metal enrichment from asymptotic giant branch (AGB) stars, SNII and SNIa
based on the metal yields from \citet{2010MNRAS.403.1413K}, \citet{2004ApJ...608..405C} and \citet{1999ApJS..125..439I}, respectively \footnote{Note that since we are tracking individual stars, the metal yields we adopt are mass-dependent rather than IMF-averaged.}.
We also include metal diffusion following  \citet{2010MNRAS.407.1581S} and \citet{2013MNRAS.434.3142A} to account for the sub-grid mixing of metals.

\subsection{IMF Sampling}\label{sec:imf_sample}

The newly formed star particles inherit their progenitor SPH particle mass $m_{\rm gas}$,
which is 1${\rm M_\odot}$ in our fiducial simulation.
The technical challenge is how to convert a group of star particles with a uniform mass distribution ($m_{\rm gas} = 1 {\rm M_\odot}$) into particles that follow an IMF while keeping the total mass unchanged.
\citet{2017MNRAS.471.2151H} introduced a method where star particles represent individual stars that collectively follow an assumed IMF,
while when resolution is poor ($m_{\rm gas} \gtrsim 100 {\rm M_\odot}$) they represent stellar populations with nonuniformly sampled IMFs.
In this work,
we revise the method by adding a distance constraint of mass transfer between star particles which we describe as follows.

The strategy of our IMF sampling is similar to \citet{2017MNRAS.471.2151H}:
we draw an array of stellar masses $m^{\rm IMF}_j$ from a given IMF and assign them to a star particle $a$ with mass $m^*_a$ until $M_{\rm IMF} \equiv \Sigma_j m^{\rm IMF}_j \geqq m^*_a$.
In practice,
$M_{\rm IMF}$ will almost always exceeds $m^*_a$.
In \citet{2017MNRAS.471.2151H},
this last sample is still kept while the residual mass $\Delta m \equiv M_{\rm IMF} - m^*_a$ will be ``borrowed'' from the next newly formed star particle $b$, i.e., $\Delta m $ is subtracted from the mass of the star particle $b$.
After the sampling,
particle masses will be adjusted such that $m^*_a \gets M_{\rm IMF}$ and $m^*_b \gets (m^*_b - \Delta m)$.
While this method ensures that the total mass of all star particles is conserved after the sampling,
it implies a mass transfer between star particles that could in principle be spatially well-separated.

In this work, we add a distance constraint such that the mass transfer can only happen within a predefined searching radius $r_{\rm sea}$.
We sample a Kroupa IMF with a mass range of [0.08M$_\odot$, 50M$_\odot$] until $M_{\rm IMF} \geqq m^*_a$.
If $M_{\rm IMF} > m^*_a$,
we search for another star particle $b$ that has not yet sampled its stellar masses within $r_{\rm sea}$.
If $M_{\rm IMF} \leqq m^*_a + m^*_b$ (meaning that there is enough mass in $b$ that can be transfered),
we transfer mass from $b$ to $a$ such that $m^*_a \gets M_{\rm IMF}$ and $m^*_b \gets (m^*_b - \Delta m)$.
If $M_{\rm IMF} > m^*_a + m^*_b$,
we search for yet another star particle $c$ within $r_{\rm sea}$ for the mass transfer,
and if $M_{\rm IMF} < m^*_a + m^*_b + m^*_c$,
we do the mass transfer such that $m^*_a \gets M_{\rm IMF}$, $m^*_c \gets (m^*_c + m^*_b  - \Delta m)$
and $m^*_b \gets 0$\footnote{The particles that are assigned with zero mass during the process will be removed.}.
We do so iteratively until either (\textit{i}) enough neighbors are found and we can keep the last sample, or (\textit{ii}) there are not enough neighbors to do the mass transfer and so we need to discard this sample.
We set $r_{\rm sea} = 4$pc and do the sampling every 2Myr.
Each star particle will end up having an array of stellar masses\footnote{At our fiducial resolution (1M$_\odot$), each particle will only have a single stellar mass.} which provide information of stellar lifetimes, UV luminosities and metal yields.

As discussed in \citet{2017MNRAS.471.2151H},
this method has the advantage of being very flexible.
When the resolution is poor (e.g., $m^* \sim 10^5 {\rm M_\odot}$), a star particle represents a star cluster following a fairly-sampled IMF and this method reduces to the stellar population model commonly adopted in cosmological simulations.
When $m^* \sim 100 - 1000 {\rm M_\odot}$,
star particles represent low-mass star clusters whose IMFs vary strongly from particle to particle due to the small samples,
which is more physical than assuming the IMF is still well-sampled at this scale.
When $m^* \sim 1 {\rm M_\odot}$,
star particles represent individual stars which collectively follow the assumed IMF.
It also has the advantage that the stellar dynamics will be better resolved as $m^* \approx m_{\rm gas}$,
as opposed to the sink-particle approach where typically $m^* \gg m_{\rm gas}$.

\subsection{SNIa distribution}\label{sec:dtd}
For SNIa,
we adopt a power-law delay-time distribution ${\rm DTD}(t) = 0.15 t^{-1.12}$ where $t$ is the delay time in units of Gyr.
The normalization is chosen such that $\int_{t_{min}}^{t_{max}} {\rm DTD}(t) dt = 1$. 
The minimal and maximal delay times are $t_{min}$ = 0.04Gyr  and $t_{max}$ = 21Gyr, chosen as the main sequence lifetime of the most massive and least massive SNIa progenitor stars, respectively.
For each low-mass star ($< 8 {\rm M_\odot}$) that has become a white dwarf (i.e., when its age is larger than its stellar lifetime),
its probability of exploding as an SNIa during the time [$t, t + \Delta t$] is $N_{Ia} m^{\rm IMF} f_{\rm WD}(t)^{-1}{\rm DTD}(t) \Delta t$ where $\Delta t$ is the timestep, $N_{Ia} = 1.5\times 10^{-3} {\rm M_\odot^{-1}}$ is the average number of SNIa progenitor stars per total mass of a single stellar population
and $f_{\rm WD} (t)$ is the mass fraction of white dwarfs.
We construct a table to obtain $f_{\rm WD} (t)$ as a function of $t$ for a Kroupa IMF.

\subsection{SN feedback}\label{sec:SNfeedback}

\begin{figure*}
	\centering
	\includegraphics[width=0.75\linewidth]{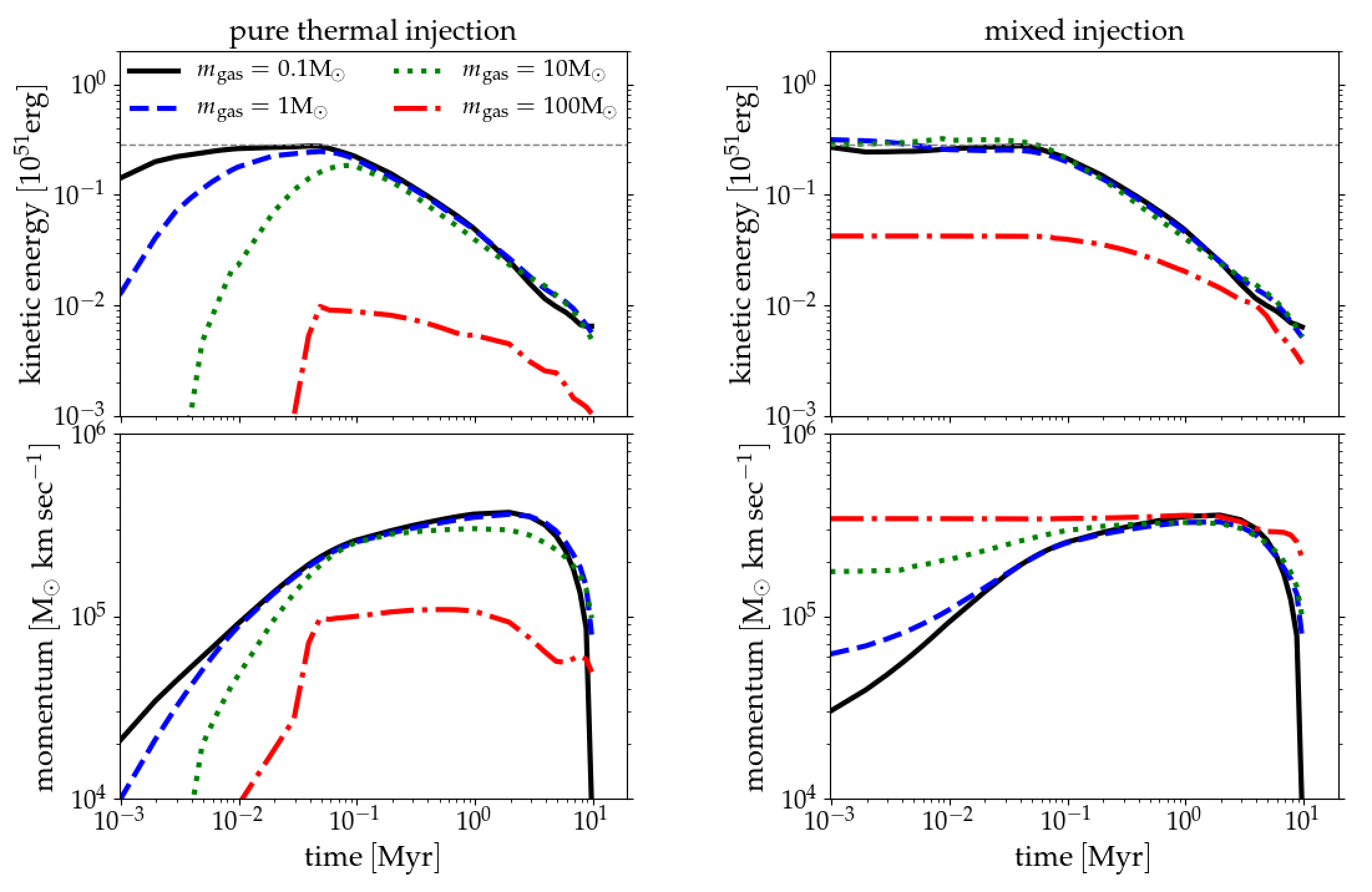}
	\caption{Time evolution of the kinetic energy (top panels) and the momentum (bottom panels) of a supernova remnant (SNR) with $E_{\rm SN} = 10^{51}$ erg in a uniform medium of $n_{\rm H} = 1$ cm$^{-3}$ at four different resolutions ($m_{\rm gas}$ = 0.1, 1, 10 and 100${\rm M_\odot}$, and the corresponding injection mass $M_{\rm inj} = 10, 100, 10^3$ and $10^4 {\rm M_\odot}$, respectively).
	The SNR with 100\% thermal injection is shown in the left panels while the SNR with the default injection scheme is shown in the right panels.
	The dashed horizontal lines indicate the expected kinetic energy in the Sedov-Taylor phase ($0.28 E_{\rm SN}$).
	When the SN becomes unresolved ($m_{\rm gas} = 100 {\rm M_\odot}$), thermal injection underestimate the terminal momentum by a factor of three,
 	while our sub-grid model by construction acquires the correct terminal momentum.}
	\label{fig:snrconv}
\end{figure*}

In contrast to \citet{2017MNRAS.471.2151H} where the energy of each SN $E_{\rm SN}$ is injected into the ISM all in the form of thermal energy,
we implement a revised energy injection scheme which we describe as follows.

\subsubsection{Exact energy partition for resolved SNe}
When an SN is resolved (see section \ref{sec:subgridSN} for definition), we inject $0.28 E_{\rm SN}$ in terms of kinetic energy while the rest $0.72 E_{\rm SN}$ is injected as thermal energy.
This energy partition is based on the exact solution of a SNR that is still in the energy conserving Sedov-Taylor (ST) phase before radiative cooling starts to kick in.
As such, the energy partition is guaranteed to be correct right after the energy injection. 
We take the canonical value $E_{\rm SN} = 10^{51}$ erg for each SN,
and inject this energy into $N_{\rm inj} = 96$ gas particles
\footnote{This value is chosen to be close to $N_{\rm ngb} = 100$, the number of particles within an SPH kernel, such that the injected region is hydrodynamically resolved.}
with a pixelwise scheme (see section \ref{sec:healpix}).

The definition of ``injecting'' kinetic energy can be ambiguous when there are strong relative motions between gas and SNe.
For example, 
gas flows converging rapidly to an SN may even be decelerated by the SN blastwave,
meaning that the gas kinetic energy actually decreases.
In practice, however,
the SN kinetic energy tends to be much higher than the kinetic energy of the gas to be injected in typical ISM conditions.
The kinetic energy injected to each particle is $e_k = 0.28 E_{\rm SN} N_{\rm inj}^{-1} $ and the corresponding velocity kick is $\Delta v = \sqrt{2e_k / m_{\rm gas}} \approx 541$ km s$^{-1} (m_{\rm gas} / {\rm M_\odot})^{-0.5}$ along the outward radial direction.
This velocity kick is much higher than the typical velocity dispersion in the ISM ($\sim 10 {\rm km s^{-1}}$) for the resolution we explore in this paper,
making it possible to have a well-defined kinetic energy injection 
\footnote{This is not the case when $m_{\rm gas} \gtrsim 100{\rm M_\odot}$.
However, at this resolution, most SNe become unresolved anyway and we would switch to the sub-grid model described in \ref{sec:subgridSN}.}.

\subsubsection{HealPix-based pixelwise injection}\label{sec:healpix}
In \citet{2017MNRAS.471.2151H}, the SN energy is injected into the nearest $N_{\rm inj}$ gas particles.
While this works fine for thermal energy injection,
it has potential issues for kinetic energy/momentum injection when the nearest gas particles are highly clumpy as it implies an anisotropic injection, i.e.,
the kinetic energy will be preferentially injected in high-density regions where most particles are located.
To address this issue,
we implement a pixel-by-pixel injection scheme using the {\sc HealPix} tessellation library \citep{2011ascl.soft07018G} and divide the $4\pi$ solid angle into $N_{\rm pix} = 12$ pixels. 
Within each pixel we look for the nearest $N_{\rm inj} / N_{\rm pix} = 8$ gas particles for the energy injection.
By doing so, we guarantee that the kinetic energy injection is isotropic at the pixel level.
Although there could still be an anisotropic particle distribution within each pixel,
this should be sufficiently small as particles are not expected to be highly clumpy at scales $N_{\rm pix}$ times smaller than the SPH kernel mass.

\subsubsection{Sub-grid model for unresolved SNe}\label{sec:subgridSN}
Both the pure thermal feedback and the one based on the exact ST energy partition become problematic when the ST phase is unresolved.
This is defined as the injection radius (typical a few resolution elements) becomes larger than the cooling radius $R_c$ where radiative cooling kicks in and the system enters a pressure-driven ``snowplow'' phase followed by a momentum-conserving phase with a terminal momentum $p_{\rm term}$.
In Lagrangian codes like ours,
the equivalent definition would be when the injection mass $M_{\rm inj} = N_{\rm inj} m_{\rm gas}$ becomes larger than the cooling mass $M_c = (4\pi/3) \rho_0 R_c^3$ where $\rho_0$ is the ambient density.
We adopt $R_c = 22.6$ pc $E_{51}^{0.29} n_{\rm H}^{-0.42}$ from \citet{2015ApJ...802...99K}
\footnote{We note that the metallicity ($Z$) in our initial setup is 10 times lower than that in \citet{2015ApJ...802...99K}. However, $R_c$ only has a very weak dependence on metallicity ($R_c \propto Z^{-0.14}$, see e.g. \citealp{1988ApJ...334..252C}).} 
where $E_{51}$ is the SN energy in units of $10^{51}$ erg and $n_{\rm H}$ is the hydrogen number density of the ambient medium in units of cm$^{-3}$.
This leads to the following requirement of resolving the ST phase: 
\begin{equation}\label{eq:coolmass}
	m_{\rm gas} < 1680 {\rm M_\odot} E_{51}^{0.87} n_{\rm H}^{-0.26} N_{\rm inj}^{-1} .
\end{equation}
When Eq. \ref{eq:coolmass} is not satisfied,
physically, a significant amount of energy has already been radiated away right after the injection,
and therefore injecting $10^{51}$ erg into $M_{\rm inj}$ leads to an incorrect evolution of the SNR.
Injecting thermal energy results in the well-known ``over-cooling'' problem which underestimates the terminal momentum as most energy would radiate away almost immediately without having much dynamical impact on the ISM.
On the other hand,
ST injection (or pure kinetic injection) overestimates the terminal momentum \citep{2017ApJ...846..133K, 2018MNRAS.477.1578H} as it wrongly assumes that the SNR still has 0.28 $E_{51}$ of kinetic energy at that scale while physically the kinetic energy has already decayed due to momentum conservation.

\clearpage

We therefore adopt a sub-grid model which has been widely adopted in the literature (see e.g. \citealp{2013ApJ...776....1K, 2014MNRAS.445..581H, 2015MNRAS.449.1057G, 2015MNRAS.450..504M, 2015ApJ...809...69S, 2015MNRAS.451.2900K, 2015MNRAS.454..238W, 2017MNRAS.466...11R, 2017arXiv170206148H, 2018ApJ...853..173K}) when the SNe are unresolved.
Instead of injecting energy,
we inject the terminal momentum 
\begin{equation}\label{eq:pterm}
	p_{\rm term} = 3.3 \times 10^5 {\rm M_\odot ~ km ~ s^{-1} } E_{51}^{0.93} n_{\rm H}^{-0.13},
\end{equation}
where the scalings are again taken from \citet{2015ApJ...802...99K} while the normalization is recalibrated by our numerical experiments (cf. Fig. \ref{fig:snrconv}).
This momentum is injected into $N_{\rm inj}$ gas particles using the same pixel-by-pixel scheme as described above.
Each particle will receive a momentum of $N_{\rm inj}^{-1} p_{\rm term}$ and the corresponding velocity kick is $\Delta v = 3.44 \times 10^3$ km s$^{-1} n_{\rm H}^{-0.13} (m_{\rm gas} / {\rm M_\odot})^{-1}$.
Following the scaling relation reported in \citet{1998ApJ...500...95T},
We also include the residual thermal energy of the SNR:
\begin{equation}\label{eq:resEth}
	E^{\rm res}_{\rm th} = 0.72 E_{\rm SN} \Big(\frac{M_{\rm inj} }{M_c}\Big)^{-2.17}.
\end{equation}
As shown in Appendix \ref{app:resEth},
this residual thermal energy has little effect on the wind properties and is included mainly for consistency.

When the injection region is highly inhomogeneous,
a single SN may be resolved along certain pixels while unresolved along the other pixels.
We therefore calculate Eq. \ref{eq:coolmass}, \ref{eq:pterm} and \ref{eq:resEth} pixel by pixel using the hydrogen number density of each pixel.

Our default SN scheme can be summarized as follows:
\begin{enumerate}
	\item When an SN is resolved (Eq. \ref{eq:coolmass} is true), inject $0.28 E_{\rm SN}$ in terms of kinetic energy and $0.72 E_{\rm SN}$ in terms of thermal energy into the neighboring particles found by the {\sc HealPix}-based algorithm. 
	\item When an SN is unresolved (Eq. \ref{eq:coolmass} is not true), inject the terminal momentum $p_{\rm term}$ and the residual thermal energy $E^{\rm res}_{\rm th}$ given by Eq. \ref{eq:pterm} and \ref{eq:resEth}, respectively. 
\end{enumerate}

\subsubsection{Numerical test}
Fig. \ref{fig:snrconv} shows the time evolution of the kinetic energy (top panels) and momentum (bottom panels) of an SNR with $E_{\rm SN} = 10^{51}$ erg in a uniform medium of $n_{\rm H} = 1$ cm$^{-3}$ at four different resolutions ($m_{\rm gas}$ = 0.1, 1, 10 and 100${\rm M_\odot}$, and the corresponding injection mass $M_{\rm inj} = 10, 100, 10^3$ and $10^4 {\rm M_\odot}$, respectively).
The SNR with pure thermal injection is shown in the left panels while the SNR with our default injection scheme is shown in the right panels.
The dashed horizontal lines indicate the expected kinetic energy in the ST phase ($0.28 E_{\rm SN}$).

In the case of thermal injection (left panels),
the time it takes to convert the thermal energy into kinetic energy to the expected amount of $0.28 E_{\rm SN}$ in the ST phase lengthens as the resolution becomes worse,
which reflect the capability of the system to converge to the exact solution and generate the right amount of momentum in the ST phase.
Once the radiative cooling kicks in,
the momentum of the SNR rapidly plateaus and eventually drops when the SN bubble falls back and closes.
According to Eq. \ref{eq:coolmass}, the SN is resolved when $m_{\rm gas} \lesssim 20 {\rm M_\odot}$.
As expected, the terminal momentum is converged ($p_{\rm term} \approx 3.3 \times 10^5 {\rm M_\odot ~ km ~ s^{-1} }$) in cases of $m_{\rm gas}$ = 0.1, 1 and 10 ${\rm M_\odot}$ while it becomes significantly underestimated in the $m_{\rm gas} = 100 {\rm M_\odot}$ case, which is a demonstration of the well-known ``over-cooling'' problem.

With the default injection scheme (right panels),
in the resolved cases ($m_{\rm gas}$ = 0.1, 1 and 10${\rm M_\odot}$),
the SNR acquires $0.28 E_{\rm SN}$ of kinetic energy by construction right from the beginning of the injection,
and it remains so for the entire ST phase.
The terminal momentum is converged and agrees well with that in the thermal injection case,
which is consistent with \citet{2012MNRAS.419..465D} and  \citet{2015ApJ...809...69S}.
In the unresolved case ($m_{\rm gas} = 100 {\rm M_\odot}$), since we put in the expected terminal momentum (Eq. \ref{eq:pterm}) by hand and this momentum has to be conserved afterwards,
we obtain the same terminal momentum as the resolved ones,
which is around a factor of 3 higher than that in the $m_{\rm gas} = 100 {\rm M_\odot}$ case with pure thermal injection.

We emphasize that putting in the expected $p_{\rm term}$ by hand for  the unresolved SNe is ultimately still a sub-grid model.
There are other aspects that this model cannot capture (for example, the low-density bubble and the dense shell),
which can play an important role in simulations.
This motivates us to test whether this sub-grid model provides better convergence compared to the pure thermal injection when applied to more realistic simulations (see section \ref{sec:SNinjectResult}).


\subsection{Initial conditions}

The initial conditions consist of an exponential gaseous disk subject to an external gravitational potential of a dark matter halo.
The halo follows the Navarro-Frenk-White profile \citep{1997ApJ...490..493N} with virial mass $M_{\rm vir} = 10^{10} {\rm M_\odot}$, virial radius $R_{\rm vir} = 44.4$kpc and concentration $c = 17$.
The choice of $c$ is motivated by the scaling relation from cosmological N-body simulations in \citet{2014MNRAS.441.3359D}.
The total mass of the disk is $M_{\rm gas} = 10^7 {\rm M_\odot}$.
The initial temperature of gas is $T = 10^4$ K, while the density profile of the disk (in a cylindrical coordinate) is 
\begin{equation}
	\rho_{\rm gas} (R,z) = \frac{M_{\rm gas}}{4\pi R^2_0 z_0} \exp(-\frac{R}{R_0}) \exp(-\frac{z}{z_0}) 
\end{equation}
where $R_0$ and $z_0$ are the scale-length and scale-height of the disk, respectively.
We set $r_0$ = 0.4 kpc such that the gas surface density at $r$ = 0 is 10 ${\rm M_\odot pc^{-2}}$.
We set $z_0$ = 0.5 kpc, which is a somewhat arbitrary choice as the scale-height of the disk changes rapidly once the simulation begins (since the gas will cool and collapse onto the midplane).
The initial gas metallicity is $Z = 0.1 Z_{\rm \odot}$ and the dust-to-gas ratio is 0.1\% in mass.
The rotation velocity of the disk follows the circular velocity of the halo including the correction of pressure gradient.
The SPH particle mass is $m_{\rm gas} = 1 {\rm M_\odot}$.
The gravitational softening length for gas is set to be the same as the SPH smoothing length.
Once a gas particle is converted into a star particle, we adopt a fixed softening length of 0.3 pc \footnote{With our resolution ($m_{\rm gas} = 1 {\rm M_\odot}$),  star particles actually represent individual stars and therefore the two-body interaction is physical rather than numerical. We still adopt a small but finite softening length just to prevent them from having too small timesteps.}.

\section{Results}\label{sec:results}

\subsection{Definitions}\label{sec:def}

Before discussing our results,
we first need to give clear definitions of wind quantities and how we measure them in the simulations.
We quantify the winds by the fluxes of mass ($F_{\rm m}$), momentum ($F_{\rm p}$) and energy ($F_{\rm e}$) as in the fluid equations,
viz.
\begin{equation}
	F_{\rm m} = \rho {\bf v} ,~~
	F_{\rm p} = \rho {\bf v}{\bf v} + P ,~~
	F_{\rm e} = (\rho e_{\rm tot} + P) {\bf v},
\end{equation}	
where $\rho$, ${\bf v}$ and $P$ are the density, velocity and pressure of gas, respectively. $e_{\rm tot} = 0.5 ||{\bf v}||^2 + u$ is the total energy (kinetic + thermal) per mass and $u$ is the specific thermal energy.
Integrating the fluxes over a measuring surface $S$,
we obtain the flow rate of mass ($\dot{M}$), momentum ($\dot{p}$)\footnote{Note that the integral of $F_{\rm p}$ is a vector. We define $\dot{p}$ by taking only the vector component normal to $S$.} and energy ($\dot{E}$) as
\begin{equation}
	\dot{M} = \int_{S} F_{\rm m} \cdot \hat{n} dA, ~~\dot{p} = \int_{S} F_{\rm p} \cdot \hat{n} dA, ~~\dot{E} = \int_{S} F_{\rm e} \cdot \hat{n} dA
\end{equation}
where $\hat{n}$ is the outward (away from the galaxy) unit normal vector of the area $dA$.
Far away from the disk, 
we define the measuring surface as a spherical shell of radius $r_s$ 
less the region where $|z| / r_s < 1 / \sqrt{2}$,
which encloses the majority of the outflowing gas and 
excludes the flaring disk component at large galactocentric radii.
On the other hand,
as the winds are mostly vertical rather than radial in the vicinity of the disk,
a spherical shell is no longer an appropriate measuring surface.
Therefore, when $r_s \leqq 1.5$kpc,  we measure the wind properties with two planes parallel to the disk's midplane at $z = \pm r_s$\footnote{This leads to a small discontinuity of the wind properties at $r_s \approx 1.5$kpc as the transition is not perfectly smooth (cf. Fig. \ref{fig:mpeloadingz}, \ref{fig:mpeloadingzphases} and \ref{fig:windprofiletimeave}).}.

The flow rate of mass and energy can be further decomposed into the outflow (where the radial velocity $v_r \equiv {\bf v}\cdot \hat{n} > 0$)  and inflow ($v_r < 0$) components such that 
\begin{equation}
	\dot{M} = \dot{M}_{out} -  \dot{M}_{in}, ~~ 	\dot{E} = \dot{E}_{out} -  \dot{E}_{in}.
\end{equation}
Note that we define the inflow rates to be positive by absorbing the minus sign in ${\bf v}\cdot \hat{n}$.
The momentum flow rate cannot be decomposed into outflow and inflow as it is always positive.

In SPH simulations,
the discretized flow rates can be expressed as
\begin{eqnarray}
	\dot{M}_{out} &=& \sum_{i, v_{r,i}>0} \dfrac{m_i v_{r,i}}{dr}, \\
	\dot{M}_{in}   &=& - \sum_{i, v_{r,i}<0} \dfrac{m_i v_{r,i}}{dr}, \\
	\dot{p}                  &=& \sum_{i} \dfrac{m_i (v^2_{r,i} + (\gamma - 1) u_i)}{dr} \label{eq:eta_p}, \\
	\dot{E}_{out}  &=& \sum_{i, v_{r,i}>0} \dfrac{m_i (v^2_i + \gamma u_i) v_{r,i}}{dr} \label{eq:eta_eout},  \\
	\dot{E}_{in}     &=& - \sum_{i, v_{r,i}<0} \dfrac{m_i (v^2_i + \gamma u_i) v_{r,i}}{dr} \label{eq:eta_ein},  
\end{eqnarray}
using the relation $P = (\gamma - 1) \rho u$.
The summations are over particles that lie within the interval $r_s \pm 0.5 dr$. 
We adopt $dr = 0.1 r_s$.

It is useful to express the flow rates normalized to the star formation activities that drive the winds.
We define the following quantities:
\begin{itemize}
	\item outflow mass loading factor  $\eta_m^{out} \equiv \dot{M}_{out} / {\rm \overline{SFR}}$ 
	\item inflow mass loading factor  $\eta_m^{in} \equiv \dot{M}_{in} / {\rm \overline{SFR}}$ 
	\item momentum loading factor $\eta_p \equiv \dot{p} / (p_{\rm ej} R_{\rm SN}) $
	\item outflow energy loading factor $\eta_e^{out} \equiv \dot{E}_{out} / (E_{\rm SN} R_{\rm SN})$
	\item inflow energy loading factor $\eta_e^{in} \equiv \dot{E}_{in} / (E_{\rm SN} R_{\rm SN})$
\end{itemize}
where $R_{\rm SN} = {\rm \overline{SFR}} / (100{\rm M_\odot}) $ is the SN rate for a Kroupa IMF,
$p_{\rm ej} = 3\times 10^4 {\rm M_\odot~km~s^{-1}}$ is the momentum carried by the ejecta of a typical SN,
and ${\rm \overline{SFR}}$ is the time-averaged star formation rate of the galaxy.
We take the time-averaged SFR over the entire simulation time excluding the initial transient phase $t \in $ [0.4 Gyr, 1.6 Gyr]\footnote{While the disk reaches a quasi-steady state after $t = 0.2$Gyr, it takes a bit longer (0.4Gyr) for the winds at $R_{\rm vir}$ to do so.} because our system reaches a quasi-steady state where the instantaneous SFR fluctuates a lot but the mean SFR remains roughly constant.

\begin{figure*}
	\centering
	\includegraphics[width=0.95\linewidth]{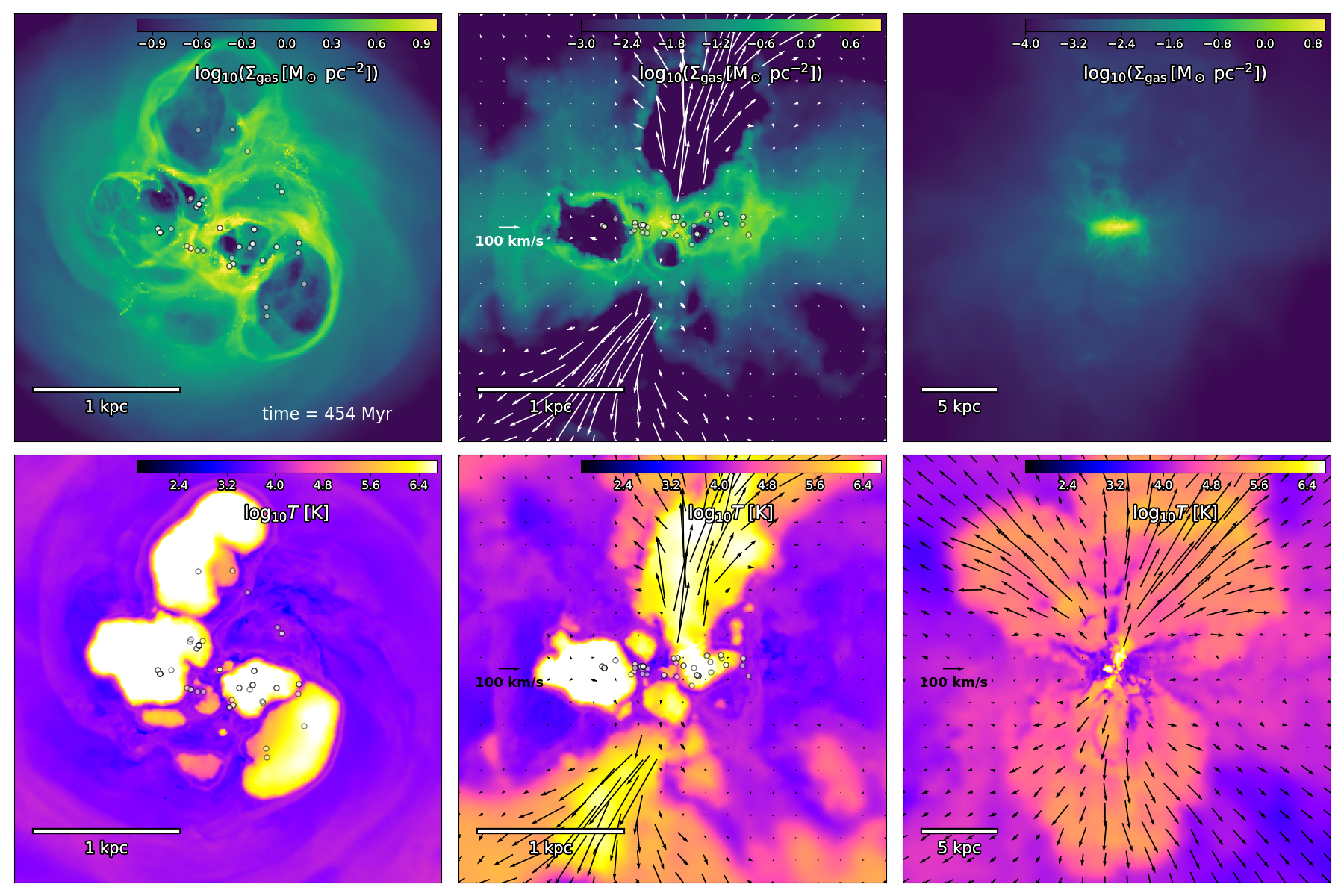}
	\caption{Maps of column density (top panels) and temperature (slices through the origin, bottom panels) of the simulated galaxy at simulation time $t$ = 454 Myr.
	The left and middle panels show the face-on and edge-on views, respectively,
	with massive stars ($> 8 {\rm M_\odot}$) over-plotted as white circles.
	The right panels show the zoom-out edge-on views.
	The fluid velocity is over-plotted as arrows (not shown in all panels).}
	\label{fig:ismfaceedgestars0399}
\end{figure*}

Likewise,
the discretized mass flow rates of metals can be defined as 
\begin{eqnarray}
	\dot{M}_{Z, out} &=& \sum_{i, v_{r,i}>0} \dfrac{Z_i m_i v_{r,i}}{dr}, \\
	\dot{M}_{Z, in}   &=& - \sum_{i, v_{r,i}<0} \dfrac{Z_i m_i v_{r,i}}{dr},
\end{eqnarray}
where $Z_i$ is the metallicity of particle $i$ (mass fraction of all heavy elements).
Correspondingly,
we define the following quantities:
\begin{itemize}
	\item outflow metal loading factor  $\eta_Z^{out} \equiv \dot{M}_{Z, out} / (m_Z R_{\rm SN})$ 
	\item inflow metal loading factor  $\eta_Z^{in} \equiv \dot{M}_{Z, in} / (m_Z R_{\rm SN})$ 
\end{itemize}
where $m_Z = 2.5 {\rm M_\odot}$ is the IMF-averaged metal mass ejected by an SN.

The metal loading factor only measures how much metals is passing through the measuring surface normalized to how much metals ejected by SNe.
A high $\dot{M}_{Z, out}$ means that a large fraction of the newly produced metals is being blown out,
but it has no information about whether metals are ``preferentially'' blown out.
To quantify the metal enrichment of the outflowing gas,
we need to define the following quantities:
\begin{itemize}
	\item outflow enrichment factor $y_Z^{out} \equiv \dot{M}_{Z, out} / (\dot{M}_{out} Z_{\rm ISM})$ 
	\item inflow enrichment factor $y_Z^{in} \equiv \dot{M}_{Z, in} / (\dot{M}_{in} Z_{\rm ISM})$ 
\end{itemize}
where $Z_{\rm ISM}$ is the time-averaged metallicity of the disk, defined as the region where $R < 1$ kpc and $|z| < 0.5$ kpc.
Note that the relation between $y_Z^{out}$ and $\eta_Z^{out}$ can be expressed as $y_Z^{out}= 0.025 ~\eta_Z^{out} / (\eta_m^{out} Z_{\rm ISM})$.

\subsection{Morphology}

In Fig. \ref{fig:ismfaceedgestars0399},
we show the maps of column density (top panels) and temperature (slices through the origin, bottom panels) of the simulated galaxy at simulation time $t$ = 454 Myr.
The left and middle panels show the face-on and edge-on views, respectively,
with massive stars ($> 8 {\rm M_\odot}$) over-plotted as white circles.
The right panels show the zoom-out edge-on views.
The fluid velocity is over-plotted as arrows (not shown in all panels).

The ISM shows a multi-phase structure which is shaped by gravity, thermal instability and SN feedback.
SNe collectively create superbubbles with sizes of a few hundred pc,
within which the gas is hot and diffuse. 
Several massive stars are located inside the bubbles.
These massive stars, when later on explode as SNe, will have a stronger dynamical impact on the ISM due to the low-density environments they reside where radiative energy losses are inefficient.
Energy input from SN feedback keeps the gaseous disk thick.
When the hot bubbles break out of the disk (i.e., when they expand well above the scale-height of the disk),
the over-pressurized hot gas ($\sim 10^6$K) vents out with velocities much higher than the warm gas ($\sim 10^4$K),
and this hot gas preferentially escapes the disk through low-density channels.
In contrast,
before the bubbles break out,
the velocities of the bubbles are much lower.
In steady state, 
the winds form a gaseous halo and exhibit a multi-phase structure.
In the zoom-out edge-on view,
temperature discontinuities can be seen where the fast moving hot winds are pushing out the warm halo gas.

\begin{figure}
	\centering
	\includegraphics[trim = 0mm 20mm 0mm 0mm, clip, width=0.9\linewidth]{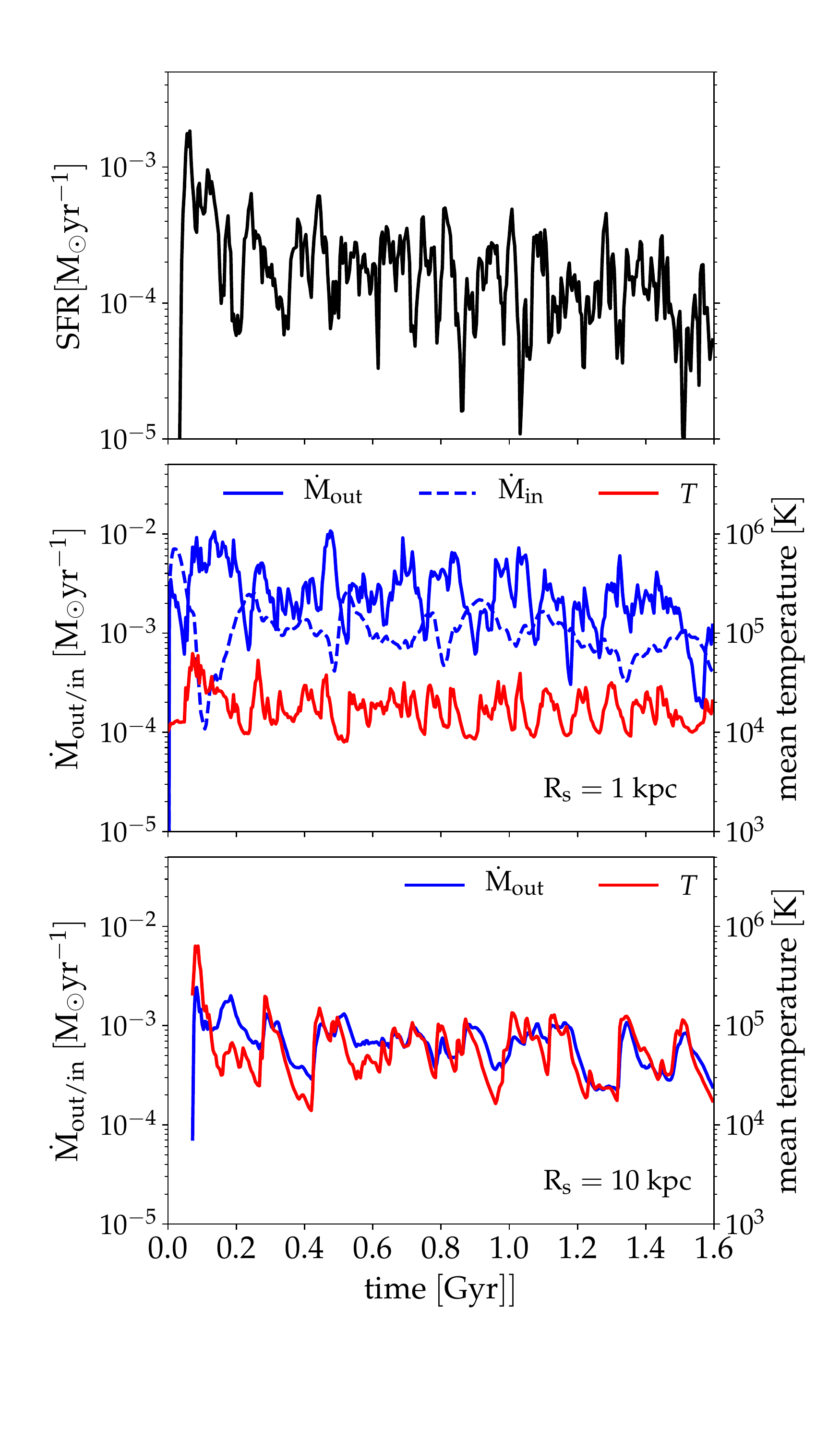}
	\caption{Top panel: time evolution of the total star formation rate (SFR). Middle panel: time evolution of the mass outflow rate ($\dot{M}_{out}$), mass inflow rate ($\dot{M}_{in}$) and mean temperature ($T$) measured at $R_{\rm s}$ = 1kpc. Bottom panel: same as the middle panel but measured at $R_{\rm s}$ = 10kpc (the inflow rate is zero). $\dot{M}_{out}$ shows a strong correlation with $T$.}
	\label{fig:ofvstime}
\end{figure}

\subsection{Time evolution}

The top panel of Fig. \ref{fig:ofvstime} shows the time evolution of the total star formation rate (SFR) of the simulated galaxy,
measured by the total stellar mass that have formed within the last 20 Myr divided by 20 Myr.
The middle panel shows the time evolution of $\dot{M}_{out}$, $\dot{M}_{in}$ and the (mass-weighted) mean temperature of gas $T$ within $r_s \pm 0.5 dr$ where $r_s$ = 1kpc.
The bottom panel is the same as the middle panel but measured at $r_s$ = 10kpc.

The SFR experiences an initial transient phase of starburst and then settles to a quasi-steady state after $t =$ 0.2 Gyr with
a mean SFR $\approx 10^{-4} {\rm M_\odot yr^{-1}}$ and strong temporal fluctuations (more than an order of magnitude). 
Although there is no cosmological inflows to replenish the gas reservoir,
the SFR can still be sustained at the same level throughout the simulation
as the galactic depletion time is much longer than the simulation time.
At $r_s$ = 1kpc,
$\dot{M}_{out}$ exhibits strong fluctuations and it anticorrelates with $\dot{M}_{in}$ but with $\dot{M}_{out} > \dot{M}_{in}$ for most of the time, meaning that there is a net mass outflow.
The mean temperature is only around $2\times 10^4$ K,
not much higher than the typical warm phase of the ISM.
At $r_s$ = 10kpc,
all the gas is flowing outward ($\dot{M}_{in} \sim 0$), and the typical timescale of variation becomes longer.
The mean temperature is higher than that at $r_s$ = 1kpc,
and there is a strong temporal correlation with $\dot{M}_{out}$:
whenever there is a rapid rise in temperature, the mass outflow rate also increases rapidly.
Note that the dynamical timescale becomes longer at large $r_s$.
Therefore, it is necessary to run the simulations for long enough time ($\gtrsim$ 1Gyr) to ensure that the entire system within the virial radius has settled to a quasi-steady state.

\subsection{Time-averaged wind properties}

\begin{figure*}
	\centering
	\includegraphics[trim = 0mm 10mm 0mm 0mm, clip, width=0.8\linewidth]{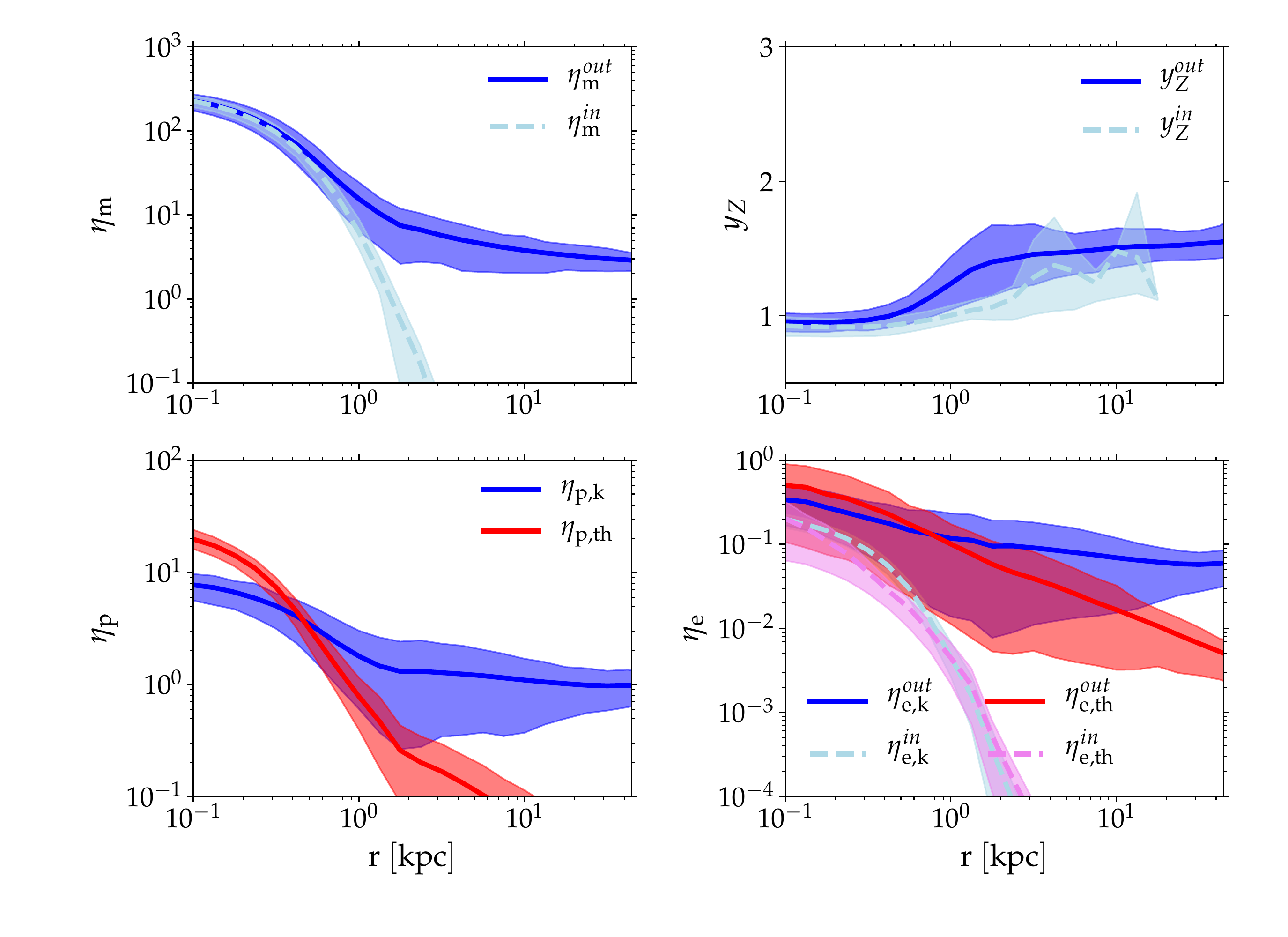}
	\caption{The mass loading factor $\eta_m$ (upper left), enrichment factor $y_Z$ (upper right), momentum loading factor $\eta_p$ (lower left) and energy loading factor $\eta_e$ (lower right) as a function of radius $r_s$ (see section \ref{sec:def} for definitions). The loading factors flatten out at $r_s > 2$kpc and remain nearly constant up to $R_{\rm vir}$, where $\eta_m^{out} \approx 3$, $\eta_p^{out} \approx 1$, $\eta_e^{out} \approx 0.05$ and $y_Z^{out} \approx 1.5$, respectively.}
	\label{fig:mpeloadingz}
\end{figure*}

In Fig. \ref{fig:mpeloadingz},
we show the time-averaged mass loading factor $\eta_m$ (upper left), metal enrichment factor $y_Z$ (upper right), momentum loading factor $\eta_p$ (lower left) and energy loading factor $\eta_e$ (lower right) as a function of measuring radius $r_s$.
$\eta_p$ and $\eta_e$ are further decomposed into kinetic (subscript ``$k$'') and thermal (subscript ``$th$'') components which correspond to the first and the second terms in Eq. \ref{eq:eta_p}, \ref{eq:eta_eout} and \ref{eq:eta_ein}, respectively.
The lines represent the time-averaged values over the interval $t \in $ [0.4 Gyr, 1.6 Gyr] while the shaded regions show the temporal fluctuations between 16 percentile and 84 percentile.

At $r_s < 2$kpc,
inflow roughly balances outflow ($\eta_m^{in} \approx \eta_m^{out}$),
indicating that the majority of the outflowing gas in this regime will eventually fall back onto the disk.
This regime is therefore dominated by the so-called ``fountain flows''.
The momentum flux at small $r_s$ is dominated by the thermal component $\eta_{p,th}$,
which indicates that the outflowing gas is mainly accelerated by the thermal pressure when it is within or close to the disk.
Similar trend can be seen in the energy fluxes,
though the difference between the thermal and kinetic components at small $r_s$ is smaller than that of the momentum flux,
mainly because the kinetic energy includes contribution from the rotation of the disk.
Unlike the mass flux where $\eta_m^{out} \approx \eta_m^{in}$,
the energy flux has a clear net outflow with both $\eta_{e,k}^{out} > \eta_{e,k}^{in}$ and $\eta_{e,th}^{out} > \eta_{e,th}^{in}$.
As will be shown later,
this is because most of the energy is carried by the hot gas, which has a net outflow.
This is contrary to the mass flux which has almost no net outflow and is dominated by the warm gas.
There is almost no metal enrichment ($y_Z^{out} \approx 1$) in this regime where the fountain flows dominate,
as a significant amount of the unenriched gas is entrained by the pressure-driven hot gas, 
diluting the enriched gas.

As $r_s$ increases,
$\eta_m^{in}$ drops rapidly towards zero and the actual winds start to emerge.
Once a gas parcel manages to reach this regime ($r_s > 2$kpc),
it will most likely escape the halo eventually,
as $\eta_m^{out}$ flattens out and remains almost constant all the way to $R_{\rm vir}$ where $\eta_m^{out} \approx$ 3.
The thermal component of the momentum loading factor$\eta_{p,th}$ drops rapidly and at large $r_s$ while the kinetic component $\eta_{p,k}$ overtakes and remains almost constant up to $R_{\rm vir}$ with $\eta_{p,k}^{out} \approx$ 1.
This suggests that ram pressure (rather than thermal pressure) becomes the main source of acceleration in this regime.
Similar trend can be seen in the energy fluxes,
where $\eta_{e,th}^{out}$ drops rapidly due to the expansion of gas while $\eta_{e,k}^{out}$ remains almost constant
up to $R_{\rm vir}$ with $\eta_{e,k}^{out} \approx 0.05$.
The majority of energy injected by the SNe is lost and only 5\% is carried with the escaping winds mainly in the form of kinetic energy.
As for metals,
$y_Z^{out}$ increases only slightly up to 1.5 at $R_{\rm vir}$,
i.e., the winds are mildly metal enriched (by around 50\%) relative to the ISM.

The fact that $\eta_m^{out}$, $\eta_p^{out}$ and $\eta_e^{out}$ all flatten out at large $r_s$ indicates that the winds are indeed in a quasi-steady state in a time-averaged sense:
for a given spherical shell, the mass, momentum and energy fluxes that flow in balance those that flow out,
which forms a steady (but not static) gaseous halo.

\subsubsection{Multi-phase winds}

\begin{figure*}
	\centering
	\includegraphics[trim = 0mm 10mm 0mm 0mm, clip, width=0.8\linewidth]{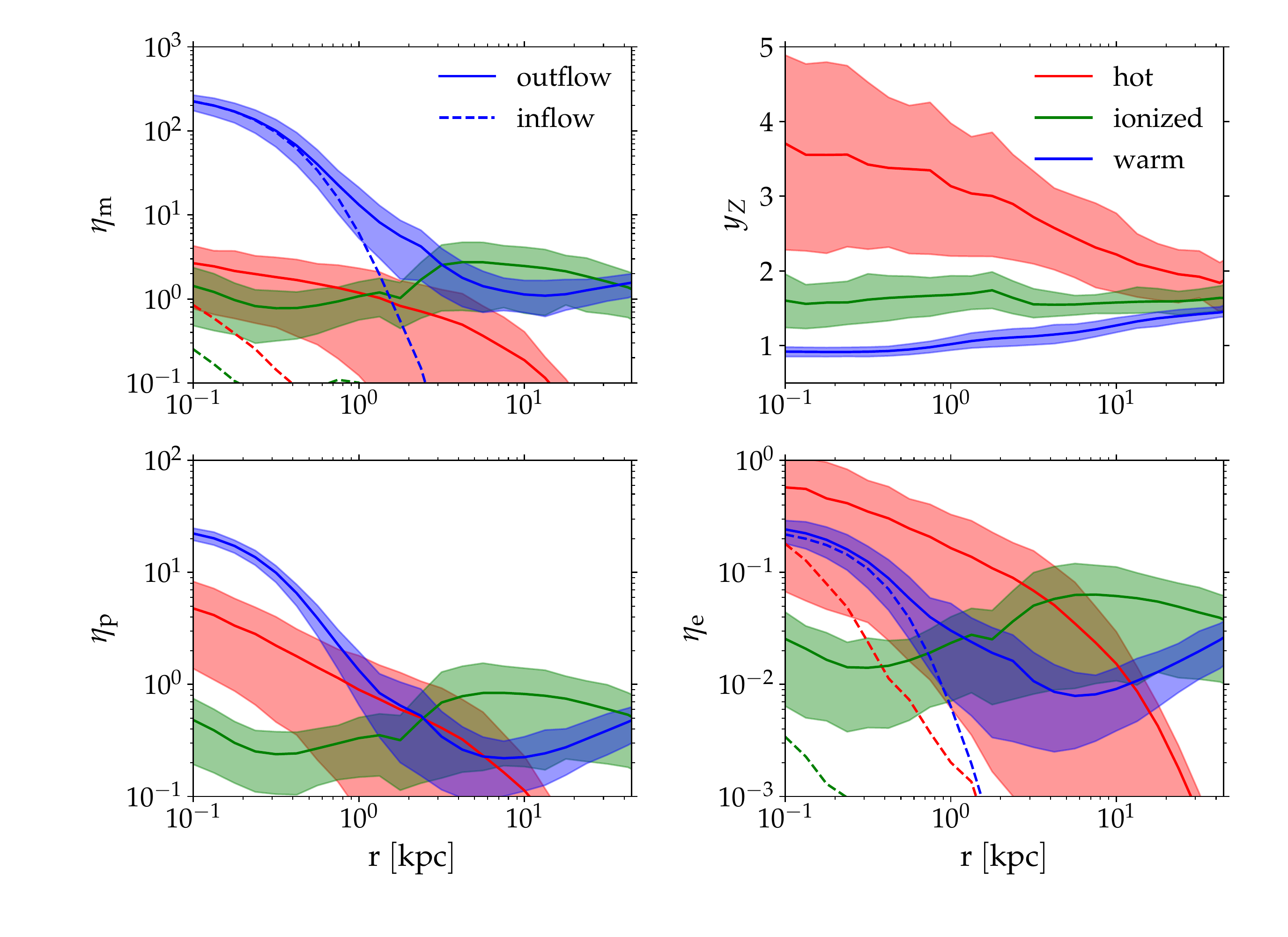}
	\caption{Same as Fig. \ref{fig:mpeloadingz} but separated into three different temperature phases: hot gas ($T > 3\times 10^5$K, in red), ionized gas ($3\times 10^4 {\rm K} < T < 3\times 10^5$K, in green) and warm gas ($T < 3\times 10^4$K, in blue). While most of the mass and momentum is carried by the warm gas, most of the energy is carried by the hot gas.}
	\label{fig:mpeloadingzphases}
\end{figure*}

In Fig. \ref{fig:mpeloadingzphases}, we show the same quantities as Fig. \ref{fig:mpeloadingz} but separate the gas into three different temperature phases:
hot gas ($T > 3\times 10^5$K, in red), ionized gas ($3\times 10^4 {\rm K} < T < 3\times 10^5$K, in green) and warm gas ($T < 3\times 10^4$K, in blue)\footnote{The warm gas here actually also includes what is conventionally called the cold ($T < 300$K) gas. However, the fraction of cold gas is negligible here. The majority of the warm gas has $T \approx 10^4$K.}.
At $r_s < 2$kpc,
the warm gas dominates both inflow and outflow which corresponds to the fountain flows.
The hot gas occupies only a small mass fraction of the flows and has a net outflow of $\eta_m^{out} \sim 1$ at $r_s = 2$kpc.
The ionized gas shows similar characteristics to the hot gas in this regime.
These two gas populations provide the necessary thermal pressure to launch the winds.
While most of the mass and momentum is carried by the warm gas,
most of the energy is carried by the hot gas.

Note that the three different phases do not evolve independently.
As $r_s$ increases,
the fraction of hot gas in the winds gradually decreases while the fraction of ionized gas increases.
This is not because the hot gas is slowing down as it moves outwards.
Instead, it is because the temperature of the hot gas decreases drops below $3\times 10^5$K.
Cooling is mainly due to the expansion of gas rather than radiative cooling,
as the latter is very inefficient at large $r_s$ where the density is low.
Similar phase transition occurs between the ionized gas and warm gas at larger $r_s$.
On the other hand,
some of the warm gas is accelerated by the ram pressure of the fast-moving hot gas.
When it gets shock-heated, it evolves from the warm phase to the ionized phase.
As a result,
in the ``halo'' regime ($r_s > 2$kpc),
the ionized gas turns out to be the dominant phase.

At small $r_s$, the warm gas is unenriched while the hot and ionized gas are metal enriched by a factor of around 3.6 and 1.6, respectively. 
Though significantly metal enriched,
as the hot gas only occupies a small mass fraction of the flows,
the overall fountain flows are close to unenriched due to the dilution by the warm gas (cf. Fig. \ref{fig:mpeloadingz}).
As $r_s$ increases,
the metallicity of different phases gradually mix with each other.
When the winds emerge ($r_s \gtrsim 2$kpc),
the hot gas has around three times higher metallicity than the warm phase.
As the winds reach $R_{\rm vir}$,
all three phases are well-mixed with $y_Z \approx 1.5$.

\subsubsection{Radial profile of winds}

\begin{figure}
	\centering
	\includegraphics[trim = 0mm 0mm 0mm 0mm, clip, width=0.8\linewidth]{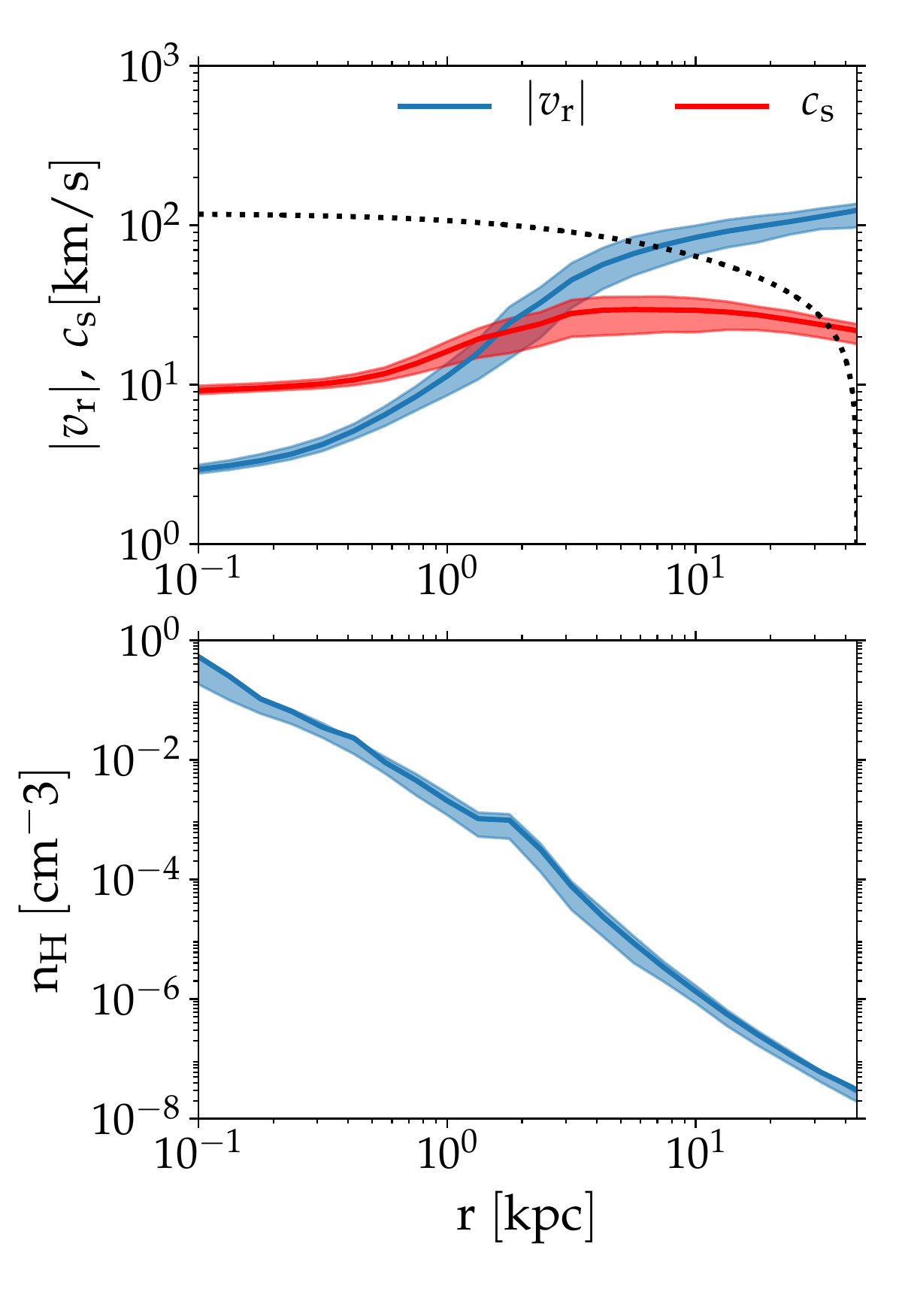}
	\caption{
		Radial profile of the winds (gas properties as a function of $r_s$).
		Upper panel: 
		radial velocity $v_{\rm r}$ and sound speed $c_s$;
		lower panel:
		hydrogen number density $n_{\rm H}$.
		All quantities are (mass-weighted) spatially- and temporally-averaged.
		The shaded areas show the temporal fluctuations between 16 percentile and 84 percentile.
		The escape velocity is overplotted in the upper panel. 
		Winds do not expand adiabatically. A significant amount of energy is added to the winds as they travel towards $R_{\rm vir}$.}
	\label{fig:windprofiletimeave}
\end{figure}

As the system reaches a steady state,
the winds form a steady but not static gaseous halo.
In Fig. \ref{fig:windprofiletimeave},
we show the radial velocity $v_{\rm r}$ and the sound speed $c_s$ (upper panel)
and the hydrogen number density $n_{\rm H}$ (lower panel) of gas as a function of $r_s$.
All quantities are (mass-weighted) spatially- and temporally-averaged.
The shaded areas show the temporal fluctuations between 16 percentile and 84 percentile.
The escape velocity, 
defined as
\begin{equation}
	v_{\rm esc}(r) = \sqrt{2 (\Phi(R_{\rm vir}) - \Phi(r))}
\end{equation}
where $\Phi(r)$ is the gravitational potential of the NFW halo,
is overplotted in the upper panel.

The mean radial velocity increases monotonically with $r_s$ from 3 km s$^{-1}$ at $r_s$ = 0.1kpc to 100 km s$^{-1}$ at $R_{\rm vir}$.
Winds are gradually accelerated as $r_s$ increases,
and only at $r_s > 7$kpc do they acquire enough kinetic energy to escape the halo ($v_{\rm r} > v_{\rm esc}$).
At $r_s < 2$kpc, the low level of $v_{\rm r}$ is mainly due to the fountain flows which move much slower than the actual winds.
At $r_s > 2$kpc where the winds emerge,
the fact that $v_{\rm r}$ increases with $r_s$ may seem counterintuitive.
As will be shown shortly,
the acceleration is due to the hot gas which acquires very high velocities ($> 100$ km s$^{-1}$) already at small $r_s$.
On the other hand, $c_s$ shows a much flatter profile compared to the monotonically increasing $v_{\rm r}$,
which leads to a transition of winds from subsonic ($r_s < 2$kpc) to supersonic ($r_s > 2$kpc) regime.
$n_{\rm H}$ decreases monotonically and scales roughly as $r^{-3}$ due to the expansion of gas,
Comparing the profiles of both $c_s$ and $n_{\rm H}$,
it is clear that the winds are not expanding adiabatically. 
A significant amount of energy is added to the winds as they travel towards $R_{\rm vir}$.

\subsection{Interaction between warm gas and hot gas}

\begin{figure*}
	\centering
	\includegraphics[trim = 25mm 20mm 10mm 5mm, clip, width=0.7\linewidth]{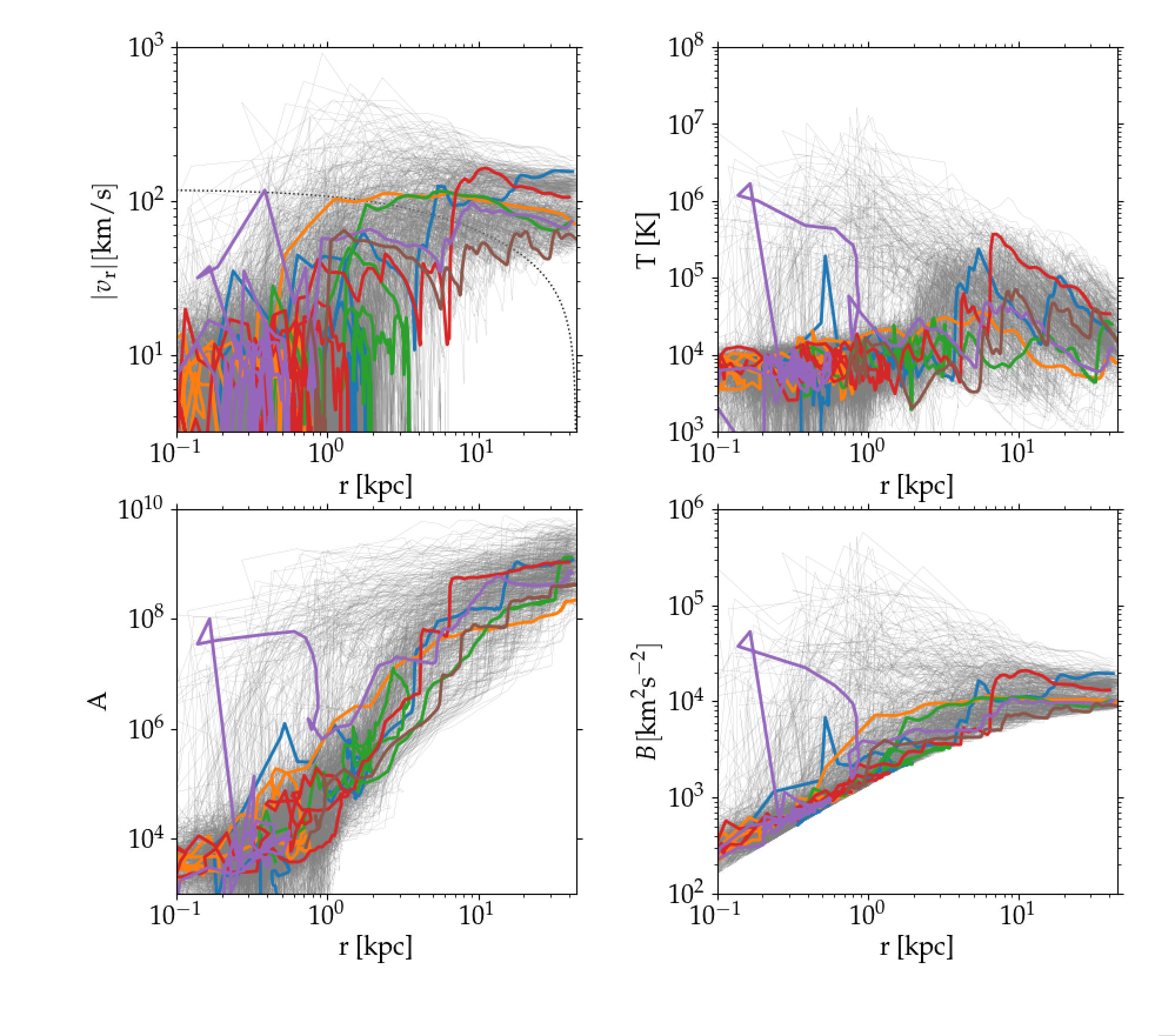}
	\caption{History of 500 randomly selected gas particles that lie within the interval [$0.95 R_{\rm vir}$, $1.05 R_{\rm vir}$] at $t = 1.6$ Gyr. 
	Four different particle quantities are shown as a function of $r_s$: $v_{\rm r}$ (top left), $T$ (top right), $A$ (bottom left) and $B$ (bottom right),
	with each thin gray trajectory represents the history of each gas particle back-traced from $t = 1.6$Gyr.
	$A$ is the entropy of gas defined as $A \equiv P / \rho^{\gamma}$,
	and $B$ is the Bernoulli parameter (see text for definition).
	The escape velocity is overplotted (dotted line) in the top left panel.
	Five particles which have their maximum temperatures lower than $3\times 10^4$K within the interval r $\in$ [1,2] kpc are highlighted in colors and in thick trajectories. Warm gas (defined at $r_s \in$ [1,2] kpc) needs to be accelerated by the ram pressure of subsequent hot winds in order to escape the halo.}
	\label{fig:historywarm}
\end{figure*}

In this section,
we investigate how winds actually escape the dark matter halo by backtracking the history of the escaping gas particles,
taking advantages of the Lagrangian nature of our code.
In Fig. \ref{fig:historywarm}, 
we show the history of 500 randomly selected gas particles that lie within the interval $r_s \in$ [$0.95 R_{\rm vir}$, $1.05 R_{\rm vir}$] at $t = 1.6$ Gyr. 
Four different particle quantities are shown as a function of $r_s$: $v_{\rm r}$ (top left), $T$ (top right), $A$ (bottom left) and $B$ (bottom right),
Each thin gray trajectory represents the history of each gas particle back-traced from $t = 1.6$Gyr.
$A$ is the entropy of gas defined as $A \equiv P / \rho^{\gamma}$,
and $B$ is the Bernoulli parameter defined as\footnote{Following \citet{2018ApJ...853..173K}, we subtract out $\Phi(0)$ in the definition of $B$ such that the gravitational term ($\Phi(r) - \Phi(0)$) is always positive.} 
\begin{equation}
	B = 0.5 v^2 + \gamma u + \Phi(r) - \Phi(0).
\end{equation} 
The escape velocity is overplotted (dotted line) in the top left panel.
Five particles which have their maximum temperatures lower than $3\times 10^4$K within the interval $r_s \in$ [1,2] kpc (the ``launching radius'') are highlighted in color with thicker trajectories.
These highlighted particles are launched as warm gas, 
and their velocities are typically lower than the escape velocity when they are launched and then slowed down by the gravitational pull of the halo.
Without further acceleration,
they will not be able to escape the halo. 
Because these particles are already supersonic when launched, there is not much thermal energy is available for the acceleration from adiabatic expansion.
However, many of them experience multiple episodes of abrupt acceleration subsequently,
which eventually give them enough kinetic energy to escape the halo.
The acceleration episodes typically correspond to a sudden rise of both temperature and entropy,
which is indicative of shocks. 
This explains why the mean sound speed in Fig. \ref{fig:windprofiletimeave} does not drop (or even increase) as what may be expected from adiabatic expansion.
In between the shock episodes,
gas is decelerated by gravity and cooled nearly adiabatically (with $A$ being constant) due to expansion.
The Bernoulli parameter, instead of staying constant, increases with $r_s$ 
suggesting that these particles keep acquiring energy as they travel towards $R_{\rm vir}$.
Interestingly,
one of the highlighted particle (in purple) reaches $10^6$K at $r_s = 0.2$kpc but then rapidly cooled back to the warm phase.
This is an example of an SN bubble that failed to break out of the disk before radiating away its thermal energy.

\begin{figure*}
	\centering
	\includegraphics[trim = 25mm 20mm 10mm 5mm, clip, width=0.7\linewidth]{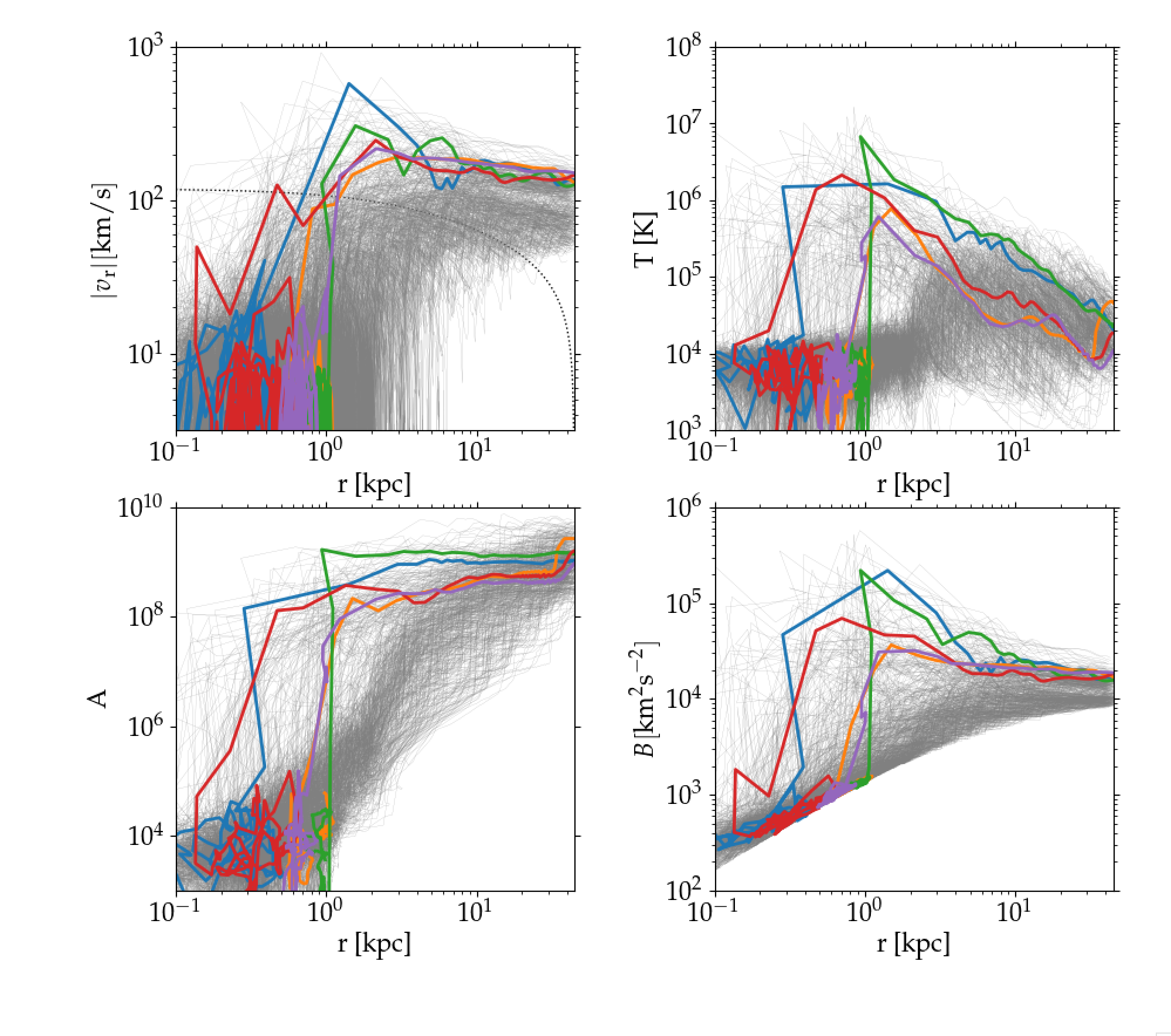}
	\caption{Same as Fig. \ref{fig:historywarm} but highlighting five particles which have their minimum temperatures higher than $3\times 10^5$K within the interval r $\in$ [1,2] kpc. Hot gas acquires velocities higher than the escape velocity right after it is launched and expands almost adiabatically afterwards.}
	\label{fig:historyhot}
\end{figure*}

We now turn to the gas particles that were hot when launched.
Fig. \ref{fig:historyhot} shows the same quantities as Fig. \ref{fig:historywarm} but highlights five particles which have their minimum temperatures higher than $3\times 10^5$K within the interval r $\in$ [1,2] kpc.
These particles correspond to the hot gas vented out from the superbubbles as they break out of the disk.
Their velocities are already higher than $v_{\rm esc}$ right after they were launched,
and remain constant or slightly decrease afterwards.
They expand and cool almost adiabatically with their entropies remain constant. 
Their Bernoulli parameters are mainly dominated by the kinetic component.
These particles could easily escape the halo without any further interaction.
Nevertheless,
they sweep up the preexisting warm gas particles in the halo (those that do not have enough energy to escape on their own), feeding energy and momentum to the latter, and both manage to escape eventually.
The winds are therefore composed of two populations of gas with different launching temperatures.
Note that as the winds travel to large $r_s$,
because the hot gas cools adiabatically and the warm gets shock-heated,
the temperature difference between these two phases decreases and eventually the winds become single-phase.

\subsection{Convergence and SN injection schemes}\label{sec:conv}

In this section,
we investigate the convergence properties by running the simulations at four different resolutions where
$m_{\rm gas}  / {\rm M_\odot}$ = 1, 5, 25 and 125 (with the corresponding injection mass $M_{\rm inj}  / {\rm M_\odot}$ = 100, 500, 2500 and $1.25\times 10^4 {\rm M_\odot}$).
The gravitational softening of stars is set to be equal to the SPH smoothing length at the typical densities where star formation occurs (recall that our star formation threshold is resolution dependent),
which turn out to be 0.2, 0.8, 4 and 10 pc, from the highest resolution run to the lowest, respectively.

\begin{figure}
	\centering
	\includegraphics[width=1\linewidth]{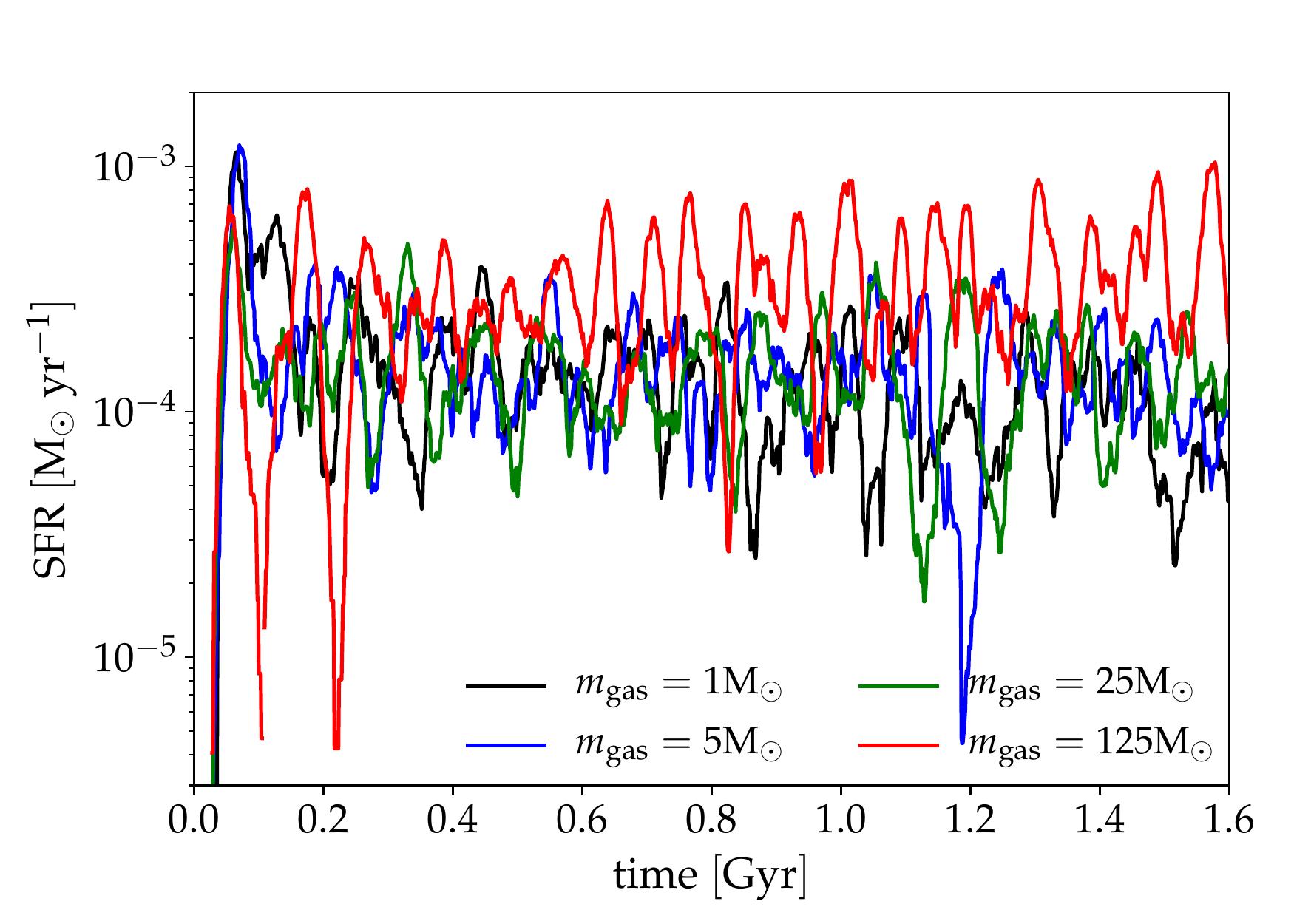}
	\caption{Time evolution of the total SFR at four different resolutions. The SFR converges at $m_{\rm gas} = 25 {\rm M_\odot}$ while it becomes slightly higher and more bursty at $m_{\rm gas} = 125 {\rm M_\odot}$.}
	\label{fig:sfrconv}
\end{figure}

In Fig. \ref{fig:sfrconv}, we show the time evolution of the total SFR at four different resolutions.
The time-averaged SFR in units of ${\rm M_\odot yr^{-1}}$ are,
from the highest resolution run to the lowest,
$1.3\times 10^{-4}$, $1.5\times 10^{-4}$, $1.5\times 10^{-4}$ and $3.6\times 10^{-4}$, respectively.
The SFR converges at $m_{\rm gas}  = 25 {\rm M_\odot}$,
while in the 125${\rm M_\odot}$-run the time-averaged SFR is higher by a factor of 2.5 and is also more bursty than the other three runs.

\begin{figure}
	\centering
	\includegraphics[width=0.9\linewidth]{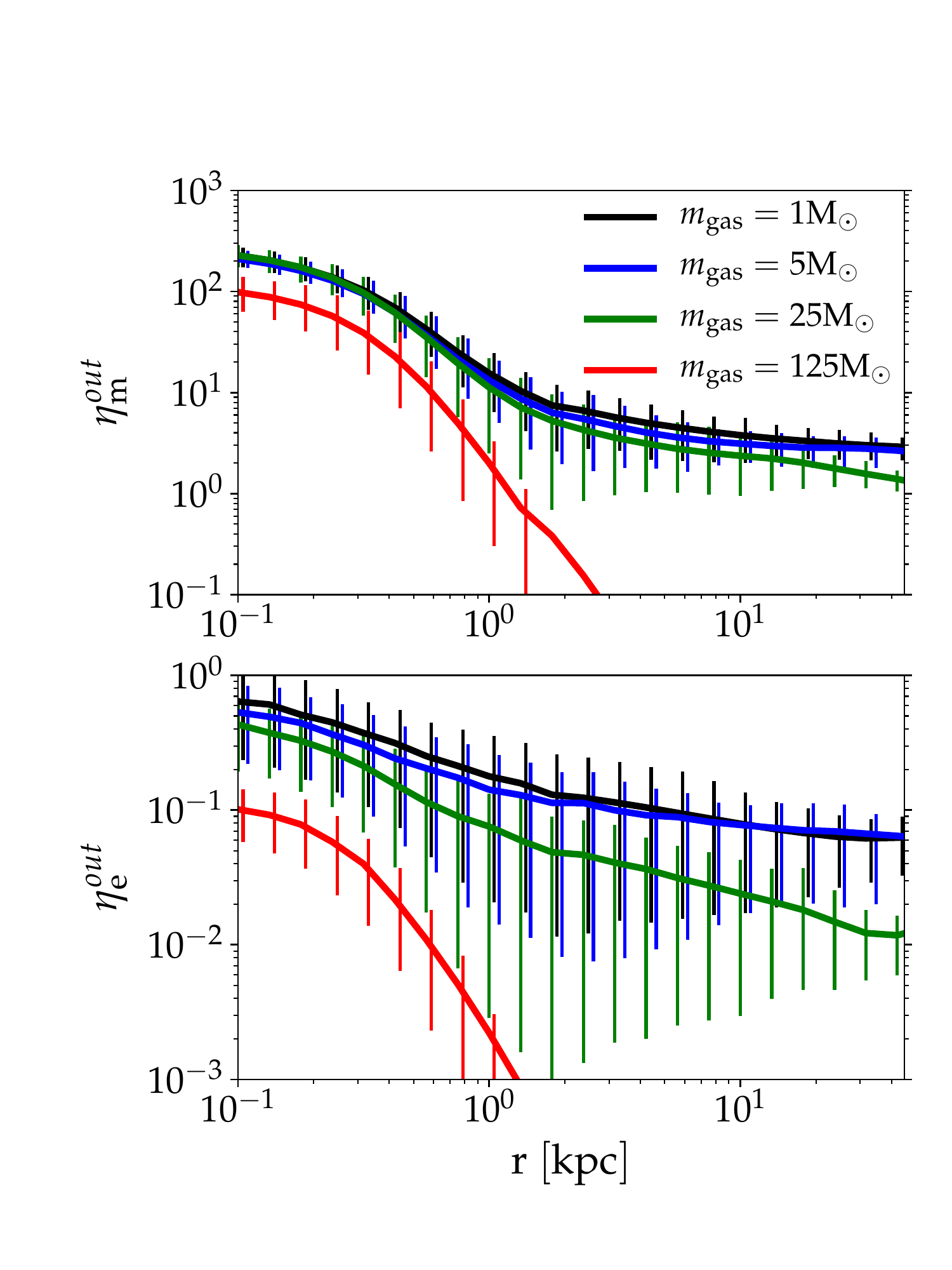}
	\caption{$\eta^{out}_m$ (top panel) and $\eta^{out}_e$ (bottom panel) as a function of $r_s$ at four different resolutions.
		The vertical error bars indicate the temporal fluctuations between 16 and 84 percentiles. 
		Wind properties converge at $m_{\rm gas} = 5 {\rm M_\odot}$ ($M_{\rm inj} = 500 {\rm M_\odot}$) where individual SNe are properly resolved.
		Our 125M$_\odot$-run fails to launch winds.}
	\label{fig:mloadconv}
\end{figure}

In Fig. \ref{fig:mloadconv}, we show $\eta^{out}_m$ (top panel) and $\eta^{out}_e$ (bottom panel) as a function of $r_s$ at four different resolutions.
The vertical error bars indicate the temporal fluctuations between 16 and 84 percentiles.
Both $\eta^{out}_m$ and $\eta^{out}_e$ converge at $m_{\rm gas} = 5 {\rm M_\odot}$ in the entire range of $r_s \in [0.1$kpc, $R_{\rm vir}]$.
The $25 {\rm M_\odot}$-run is marginally converged,
where $\eta^{out}_m$ and $\eta^{out}_e$ are lower than the converged values at $R_{\rm vir}$ by a factor of 2 and 5, respectively.
In the $125 {\rm M_\odot}$-run, however, 
both $\eta^{out}_m$ and $\eta^{out}_e$ vanish rapidly at $r_s > 2$kpc,
indicating the failure of launching winds despite its slightly higher and more bursty SFR.
Winds are more sensitive to resolutions and are more difficult to converge compared to the SFR.

\begin{figure*}
	\centering
	\includegraphics[width=0.99\linewidth]{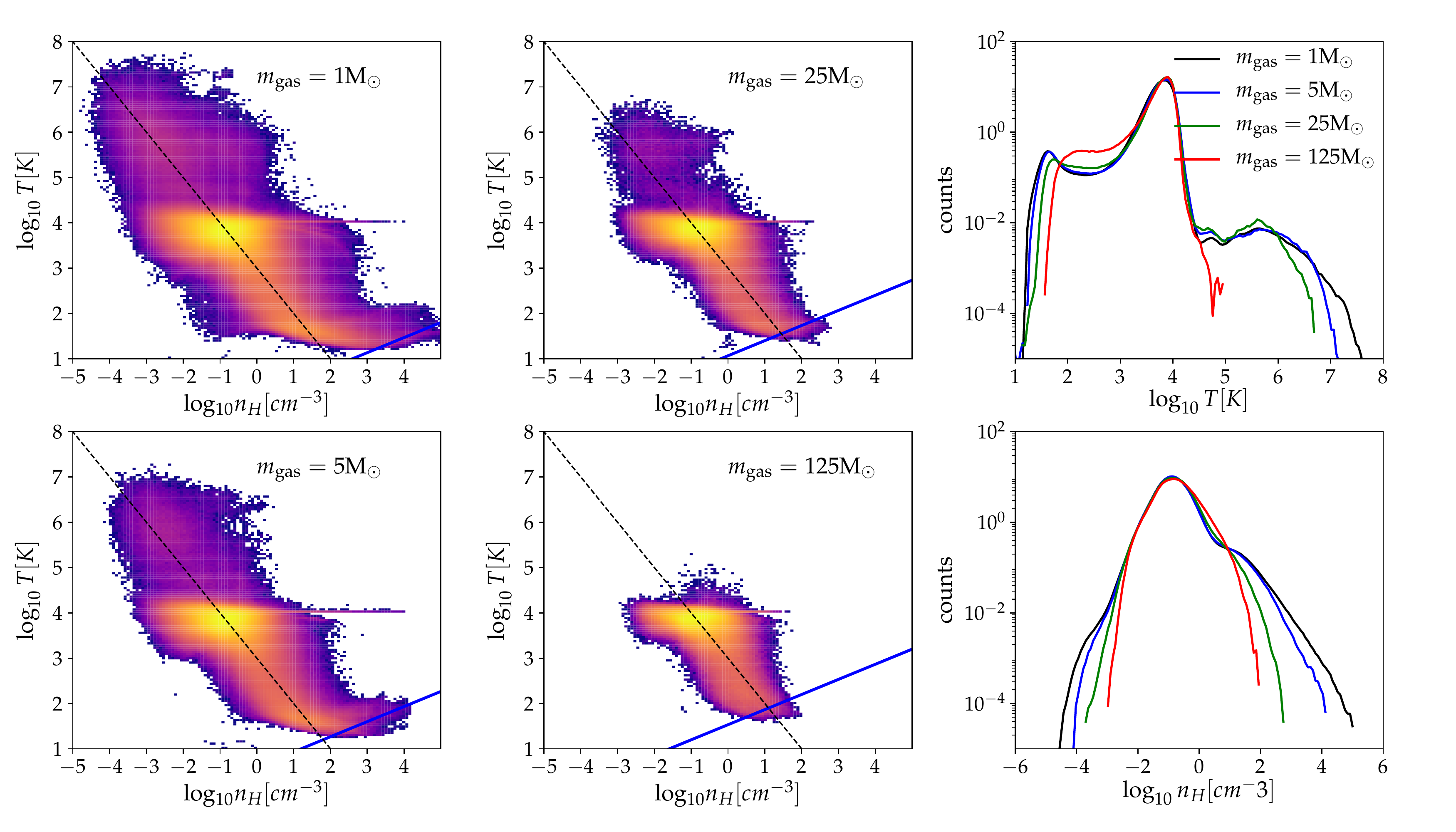}
	\caption{Time-averaged phase diagram ($n_{\rm H}$ vs. $T$) in runs with $m_{\rm gas}$ = 1M$_\odot$ (upper left), 5M$_\odot$ (lower left), 25M$_\odot$ (upper middle) and 125M$_\odot$ (lower middle), respectively.
	The blue solid line on the lower-left corner of each phase diagram indicates the (resolution -dependent) star formation threshold.
	The dashed black line indicates a constant pressure of $10^3$Kcm$^{-3}$.
	The bottom panels show the time-averaged probability density function (PDF) of $T$ (upper right) and $n_{\rm H}$ (lower right) at four different resolutions.
	Different phases of the ISM are roughly in pressure equilibrium with $P \sim 10^3 {\rm K cm^{-3}}$.
	The 125M$_\odot$-run does not generate any hot gas and therefore can not launch winds.}
	\label{fig:pd2by2converge}
\end{figure*}

In Fig. \ref{fig:pd2by2converge}, we show the time-averaged phase diagram ($n_{\rm H}$ vs. $T$) in runs with $m_{\rm gas}$ = 1M$_\odot$ (upper left), 5M$_\odot$ (lower left), 25M$_\odot$ (upper middle) and 125M$_\odot$ (lower middle), respectively.
The solid blue line on the lower-left corner of each phase diagram indicates the (resolution-dependent) star formation threshold.
The dashed black line indicates a constant pressure of $10^3$Kcm$^{-3}$.
The bottom panels show the time-averaged probability density function (PDF) of $T$ (upper right) and $n_{\rm H}$ (lower right) at four different resolutions.

In our highest-resolution run ($m_{\rm gas} = 1{\rm M_\odot}$),
the ISM has a multi-phase structure and different phases are roughly in pressure equilibrium with $P \sim 10^3 {\rm K cm^{-3}}$.
Heating from SN feedback keeps the diffuse gas ($n_{\rm H} \sim 0.1-1$cm$^{-3}$) at $T = 10^4$K,
consistent with previous studies \citep{2016MNRAS.458.3528H, 2017MNRAS.471.2151H}.
The thin horizontal line at $T = 10^4$K is due to photoionization.
At $n_{\rm H} > 100 {\rm cm^{-3}}$,
cold gas enters the isothermal regime: it decouples with the ISM, reaches a much higher pressure, and undergoes gravitationally collapse. 
Star formation occurs in the densest gas where $n_{\rm H} \sim 10^3 - 10^4$ cm$^{-3}$.
This is possible thanks to the high resolution we adopt which allows us to properly follow the gravitationally collapse up such high densities,
and it also provides a physical justification to our choice of a high star formation efficiency as we resolve the Jeans mass up to the high densities where  star formation is expected to be very efficient.
Due to the resolution-dependent star formation threshold,
the highest density the gas can reach decreases as $m_{\rm gas}$ increases.
In the 125M$_\odot$-run,
the densest gas only extend to $n_{\rm H} \sim 100$ cm$^{-3}$.
While this may lead to less clustered star formation,
it is partly compensated by its highly bursty star formation history (cf. Fig. \ref{fig:sfrconv}).

The density and temperature PDFs are well-converged at $m_{\rm gas} = 5 {\rm M_\odot}$,
and start to show deviations in the 25M$_\odot$-run, especially at the high-temperature part.
The most striking feature in the 125M$_\odot$-run (which fails to launch winds),
compared to the other higher-resolution runs,
is its complete lack of hot phase.
Since winds are driven by the thermal pressure which is mostly carried by the hot gas,
the failure of wind launching is a direct consequence of having no hot gas.

\subsubsection{SN environment}\label{sec:SNenv}

\begin{figure}
	\centering
	\includegraphics[trim = 0mm 10mm 0mm 10mm, clip, width=0.9\linewidth]{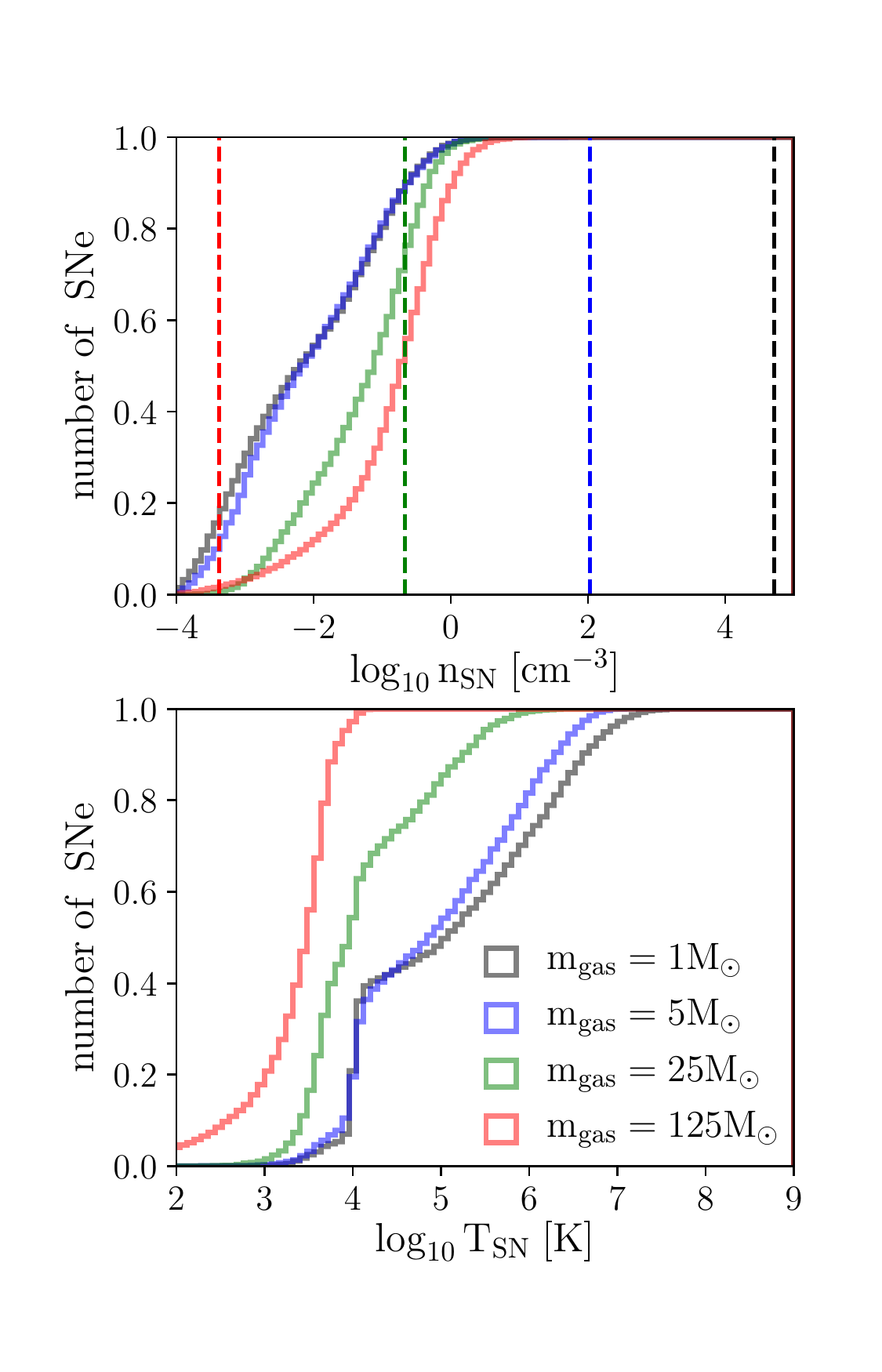}
	\caption{Cumulative distribution functions of hydrogen number density ($n_{\rm SN}$, top panel) and temperature ($T_{\rm SN}$, bottom panel) of gas where the SNe occur at four different resolutions.
	The vertical dashed lines in the top panel indicate the (resolution-dependent) critical density for which the SNe are resolved (cf. Eq. \ref{eq:coolmass}). 
	SN environment converges at $m_{\rm gas} = 5 {\rm M_\odot}$ ($M_{\rm inj} = 500 {\rm M_\odot}$). Around 60\% of SNe occur in preexisting SN bubbles that are hot and diffuse, indicating clustered SNe.}
	\label{fig:snrhot}
\end{figure}

In Fig. \ref{fig:snrhot},
we show the cumulative distribution functions of hydrogen number density ($n_{\rm SN}$, top panel) and temperature ($T_{\rm SN}$, bottom panel) of gas where the SNe occur at four different resolutions.
The vertical dashed lines in the top panel indicate the (resolution-dependent) critical density for which the SNe are resolved (cf. Eq. \ref{eq:coolmass}).
All SNe are properly resolved in the 1M$_\odot$- and 5M$_\odot$-runs.
Around 20\% SNe in the 25M$_\odot$-run are unresolved, 
and almost all SNe in the 125M$_\odot$-run are unresolved.

In the highest resolution run ($m_{\rm gas} = 1 {\rm M_\odot}$),
around 60\% SNe occur in places hotter than the typical warm ISM ($10^4$K).
This indicates that these SNe occur in preexisting hot bubbles generated by previous SNe.
The hot bubbles provide low-density environments (lower than the typical ISM $n_{\rm H} \sim 0.1 ~{\rm cm}^{-3}$) where radiative losses becom inefficient, magnifying the dynamical impact of SNe.
In fact, this is already hinted qualitatively in Fig. \ref{fig:ismfaceedgestars0399} where several massive stars can be found within the hot and diffuse superbubbles.
There is almost no SN that occurs in dense star-forming gas.
This is due to the effect of photoionization which evacuates the gas prior to the SN events,
consistent with previous studies \citep{2017MNRAS.466.3293P, 2017MNRAS.471.2151H}.
The cumulative distribution functions are converged at $m_{\rm gas} = 5 {\rm M_\odot}$.
In the 25M$_\odot$-run,
SNe occur in colder and denser environments,
though the majority of them are still resolved.
In the 125M$_\odot$-run,
no SNe occur in environments where $T_{\rm SN} > 10^4$K due to the lack of hot phase (cf. Fig. \ref{fig:pd2by2converge}),
and most SNe occur in typical ISM density $n_{\rm H} \sim 0.01 - 1 ~{\rm cm}^{-3}$.
The fact that $n_{\rm SN}$ increases with $m_{\rm gas}$ makes SNe even more poorly-resolved.
As almost all SNe are unresolved,
SN feedback is mostly done by $p_{\rm term}$-injection.

\subsubsection{SN injection scheme}\label{sec:SNinjectResult}

\begin{figure}
	\centering
	\includegraphics[trim = 0mm 20mm 0mm 0mm, clip, width=0.8\linewidth]{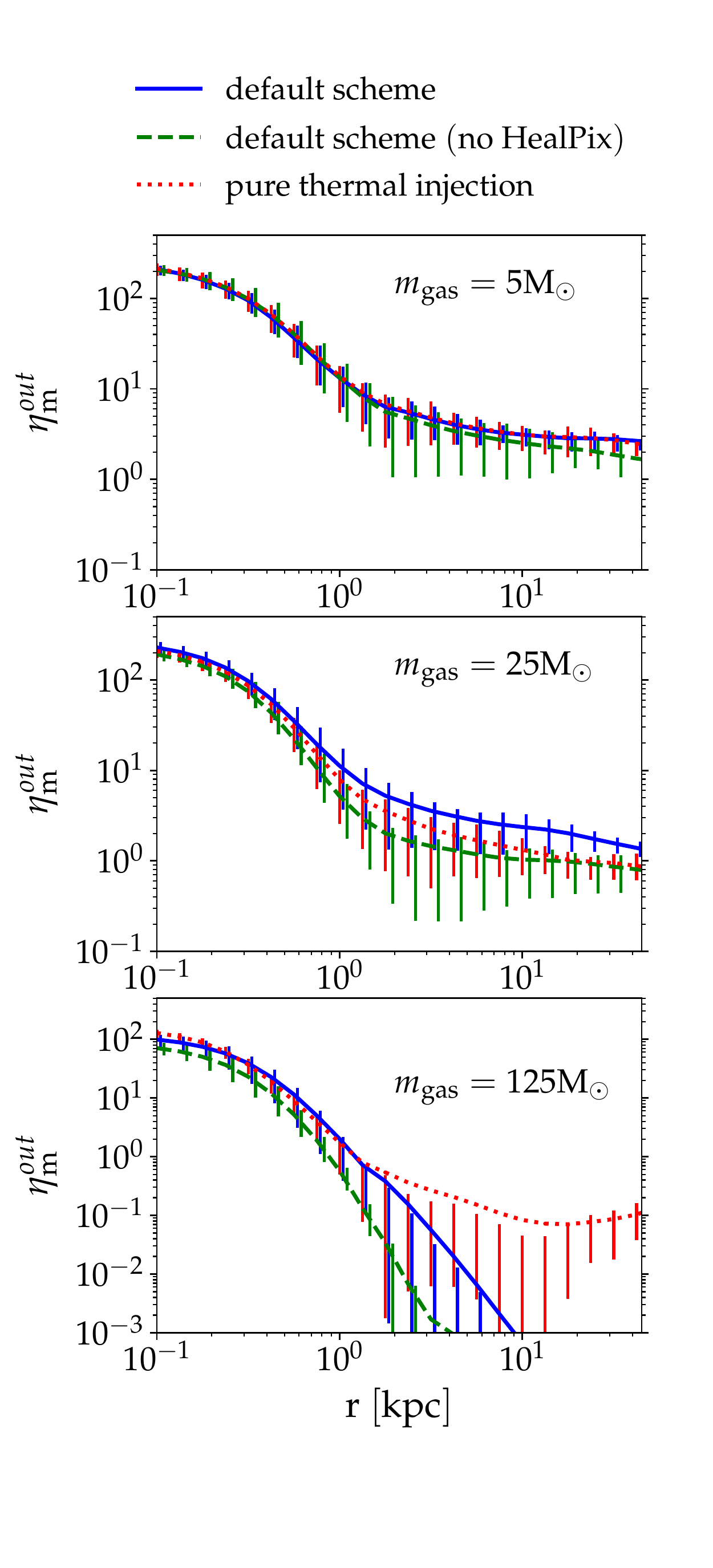}
	\caption{Outflow mass loading factor as a function of radius with three different injection schemes of SN feedback:
		(\textit{i}) default scheme (blue solid curves), (\textit{ii}) default scheme without {\sc HealPix} (green dashed curves) and (\textit{iii}) pure thermal injection (red dotted curves). See section \ref{sec:SNfeedback} for details of the default scheme. When SNe are resolved, all schemes lead to the same (converged) wind properties. When SNe are unresolved, all schemes fail to launch winds as strong as the converged ones, even with our default scheme where the SN terminal momentum is injected.}
	\label{fig:mloadinject}
\end{figure}

In this section,
we investigate how different SN injection schemes affect our results.
We compare the following three different schemes:
\begin{itemize}
	\item default scheme as described in section \ref{sec:SNfeedback}
	\item default scheme without the {\sc HealPix}-based injection
	\item pure thermal injection
\end{itemize}
In Fig. \ref{fig:mloadinject},
we show the outflow mass loading factor as a function of radius with three different injection schemes of SN feedback.
The top, middle and bottom panels are runs with $m_{\rm gas} = $5, 25 and 125 M$_\odot$, respectively.
The vertical error bars show the temporal fluctuations between 16 and 84 percentiles.

In the 5M$_\odot$-runs,
three different injection schemes lead to almost indistinguishable radial profile of $\eta_m^{out}$ in the entire range of $r_s \in$ [0.1kpc, 44.4kpc].
It has been shown that for an SNR in a uniform medium,
its evolution is insensitive to the form of injected energy (kinetic vs. thermal) as long as the SN is resolved \citep{2012MNRAS.419..465D,
2015ApJ...809...69S}.
It is reassuring that we can further extend this statement in a complicated system like our simulated galaxy:
the wind properties are insensitive to the form of injected energy as long as individual SNe are resolved,
because there is enough time for each SNR to naturally evolve towards the ST solution.
The effect of anisotropic injection is also negligible here,
presumably because the ambient density structure of SNe is properly resolved such that even without the {\sc HealPix} approach, 
the injection is still sufficiently isotropic.

As the resolution coarsens,
the injection scheme starts to make a difference.
In the 25M$_\odot$-runs,
the default injection with and without {\sc HealPix} show a factor of two difference in $\eta_m^{out}$ in the range of $r_s \in$ [2kpc, $R_{\rm vir}$].
Without {\sc HealPix},
kinetic energy is preferentially (and unphysically) injected into overdensities of gas,
which weakens the winds in an inhomogeneous medium as the overdensities are more difficult to be accelerated.
The pure thermal injection also leads to a slightly lower $\eta_m^{out}$ at $R_{\rm vir}$ (which coincidentally agrees with the default injection without {\sc HealPix}) compared to the default injection.
This suggests that the pure thermal injection may already suffer from the overcooling problem at $m_{\rm gas} = $25 M$_\odot$, 
as the SNe are only marginally resolved (cf. Fig \ref{fig:snrhot}).
We note that the differences between these runs are only a factor of two which are comparable to or even less than the temporal fluctuations.

In the 125M$_\odot$-runs,
the injection scheme has a significant effect on $\eta_m^{out}$.
Since most SNe are unresolved at this resolution, 
the pure thermal injection suffers from the overcooling problem as expected and thus underestimates $\eta_m^{out}$ by around one order of magnitude.
Surprisingly,
the default injection, which is mainly $p_{\rm term}$-injection at this resolution (cf. Fig \ref{fig:snrhot}), 
fails even more in terms of wind launching,
with $\eta_m^{out}$ three orders of magnitude lower than the converged values already at $r_s \sim 7$kpc and almost zero mass flux at $R_{\rm vir}$.
Momentum injection without {\sc HealPix} fails even more due to its mass bias,
which preferentially inject momentum into the directions parallel to the disk plane.

\begin{figure*}
	\centering
	\includegraphics[trim = 80mm 0mm 80mm 0mm, clip, width=0.6\linewidth]{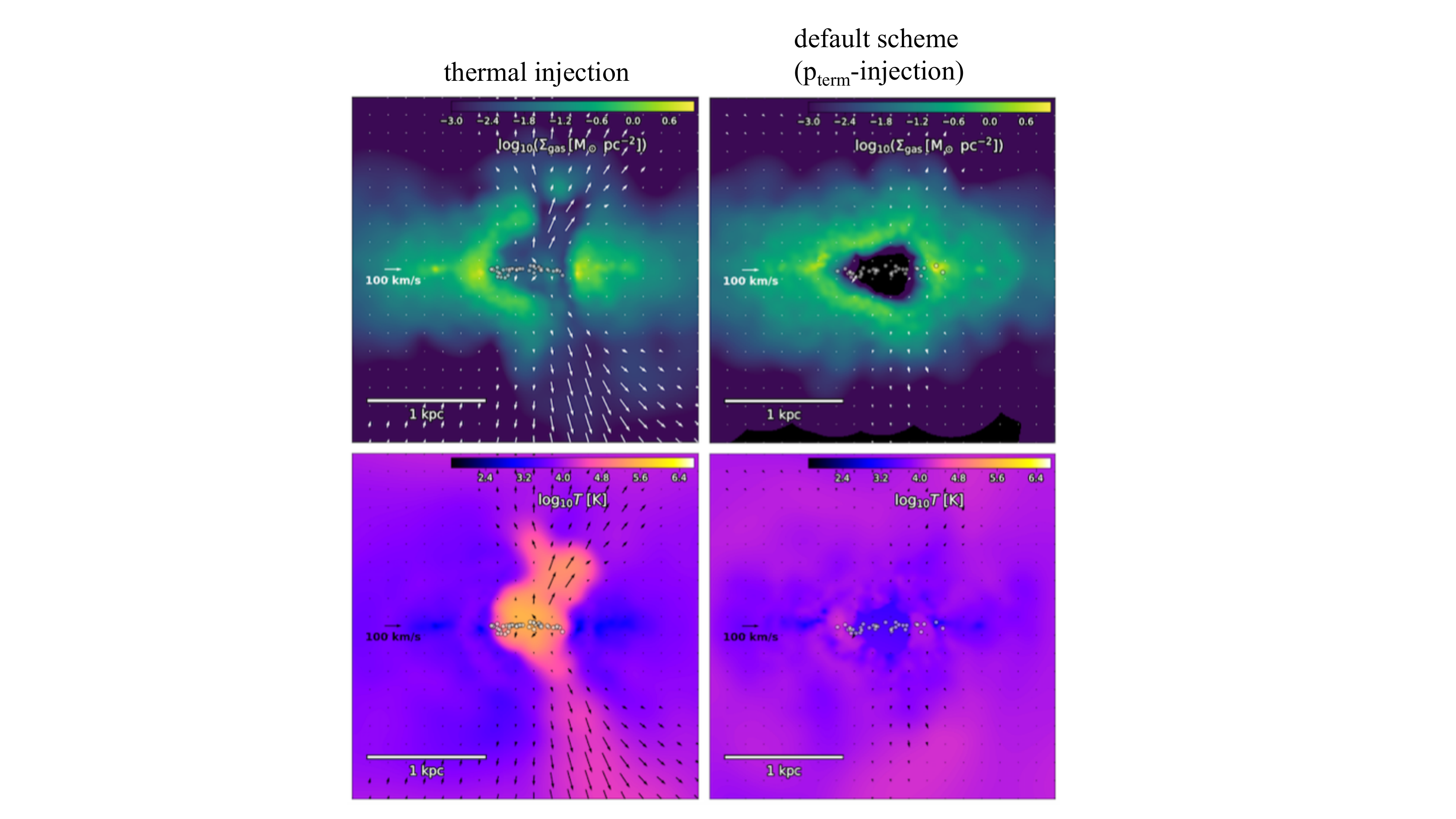}
	\caption{Maps of column density (top panels) and temperature slices (bottom panels) in the edge-on view with $m_{\rm gas} = 125 {\rm M_\odot}$ ($M_{\rm inj} = 1.25\times 10^4 {\rm M_\odot}$).
	The left panels are with pure thermal injection while the right panels are with our default scheme which is mostly $p_{\rm term}$-injection at this resolution,
	shown at the time where the most significant SN-bubbles are developed.
	Massive stars are overplotted as white circles.
	The fluid velocity is over-plotted as arrows. SN bubbles with $p_{\rm term}$-injection are filled with warm gas instead of hot gas, which makes them difficult to launch winds.}
	\label{fig:coldbubble}
\end{figure*}

The reason that $p_{\rm term}$-injection leads to even weaker winds than pure thermal injection does is due to its assumption that the thermal energy has already been radiated away right from the injection.
As such,
winds can only be momentum-driven in this case.
Apparently, it is insufficient to launch winds solely by momentum (at least in our simulations).
In Fig. \ref{fig:coldbubble}, 
we show the maps of column density (top panels) and temperature slices (bottom panels) in the edge-on view with $m_{\rm gas} = 125 {\rm M_\odot}$.
The left panels are with pure thermal injection while the right panels are with our default scheme which is mostly $p_{\rm term}$-injection at this resolution,
shown at the time where the most significant SN-bubbles are developed.
Massive stars are overplotted as white circles.
The fluid velocity is over-plotted as arrows.
The SN-bubbles in the left and right panels are similar in size, but they have strikingly different gas temperatures.
Thermal injection could generate hot bubbles in situations where SNe are highly clustered,
while $p_{\rm term}$-injection generates warm bubbles by construction even with highly clustered SNe.
By the time the bubbles break out,
only the hot bubble is able to vent out hot gas with high velocities and launch the pressure-driven winds.
We emphasize that pure thermal injection does still suffer from the overcooling problem: 
most of the time (when SNe are not clustered enough) the hot gas cools numerically and winds can not be launched as efficiently.

\section{Discussion}\label{sec:discussion}

\subsection{Comparison with previous studies}

\citet{2017ApJ...841..101L} use stratified-box simulations to study galactic winds across a range of gas surface density.
They find $\eta_m^{out} \sim 6$, $\eta_e^{out} \sim 0.2$ and $\eta_Z^{out} \sim  0.7$ in their low-surface-density model (their Fig. 10). 
This agrees surprisingly well with our model which finds $\eta_m^{out} \sim 7$, $\eta_e^{out} \sim 0.1$ and $\eta_Z^{out} \sim  0.55$ at the launching radius,
which is remarkable as their numerical model is very different from ours:
they use the grid-based code {\sc Enzo} \citep{2014ApJS..211...19B} without self-gravity and star formation and inject SNe at random locations with a predefined SN rate.

\citet{2017MNRAS.470L..39F} conduct isolated-disk simulations with the grid-based code {\sc Athena} \citep{2008ApJS..178..137S}.
They adopt a similar initial conditions to ours but their numerical model do not include self-gravity, cooling below $10^4$K and star formation,
and the SN rate is determined by the local gas free-fall time and an efficiency parameter mimicking a Schmidt-law star formation model.
Their model with lowest gas surface density ($\Sigma_{\rm gas} = 10 {\rm M_\odot}$) and individually injected SNe shows $\eta_m^{out} \sim 1$ and $\eta_e^{out} \sim 0.01$,
which are significantly smaller than ours, especially for the energy loading factor.

The fact that our results agrees better with \citet{2017ApJ...841..101L} than with \citet{2017MNRAS.470L..39F} is likely due to the SN environment.
\citet{2017ApJ...841..101L} put SNe at random locations and so many of them will occur in diffuse environments.
In our case,
as shown in Fig. \ref{fig:snrhot},
most of SNe also occur in densities lower than the peak ISM density because of SN clustering.
In contrast,
in \citet{2017MNRAS.470L..39F}, SNe preferentially occur in high-density environments,
which would lead to much weaker winds as shown in \citet{2016MNRAS.456.3432G}.
Indeed,
when they group several SNe together to mimic the effect of SN clustering,
they find loading factors much similar to ours ($\eta_m^{out} \sim 3$ and $\eta_e^{out} \sim 0.1$ in their $f_{\rm cl}=10$ model).

Interestingly,
the winds predicted by our model are much weaker than what cosmological simulations typically adopt (or effectively adopt) to reproduce realistic galaxy population (see e.g. \citealp{2011MNRAS.415...11D, 2014MNRAS.444.1518V, 2015MNRAS.446..521S}).
This suggest that either other important feedback processes are still missing in our model,
or that the wind loading factors in cosmological simulations need to be revised down.
More observational constraints on galactic winds of dwarf galaxies are highly desirable (e.g. \citealp{2018MNRAS.477.3164M}).

\citet{2018ApJ...853..173K} investigate galactic winds in a solar-neighborhood condition with stratified-box simulations and a self-consistent model that follows self-gravity, cooling, star formation and SN feedback.
They find that the warm gas forms fountain flows while the hot gas has nearly constant $\eta_m^{out}$ and $\eta_e^{out}$ above 1 kpc.
Our dwarf galaxy shows similar wind properties in the sense that the fountain flows are mostly warm and that the loading factors of winds remain nearly constant after launching ($r_s \gtrsim 2$kpc).
In addition, the typical velocity of warm gas at launching is also comparable (50 $-$ 100km s$^{-1}$).
However,
they find much lower loading factors compared to ours ($\eta_m^{out} \sim 0.1$ and $\eta_e^{out} \sim 0.02$).
In addition,
beyond the launching radius,
they do not find energy transfer from hot gas to warm gas to be significant,
while in our case a fraction of warm gas is efficiently accelerated by the hot gas and eventually escapes the halo.
The discrepancy should not be too surprising as these two set of simulations model very different galaxies.
Our dwarf galaxy has a much shallower gravitational potential
which makes warm gas easier to be accelerated to large $r_s$.
As it moves outward and expands,
its cross section increases,
making it even easier to be accelerated by ram pressure.


\subsection{Isotropic momentum injection: when does it matter?}

\citet{2018MNRAS.477.1578H} have introduced a method of injecting SN momentum isotropically by carefully distributing momentum based on the effective areas of neighboring particles.
Their method has the advantage that isotropy is imposed on a particle level,
as opposed to our method where there can still be anisotropy within each pixel (though the number of pixels can easily be increased if desired).
However, since their method still relies on a single neighbor search to find the $N_{\rm inj}$ nearest neighbors,
in extreme situations where all nearest neighbors are locates within a small solid angle of an SN (e.g. a dense clump),
the injection can still be anisotropic.
Our {\sc HealPix}-based method has the advantage that isotropy is guaranteed on a pixel level (i.e. there will be no empty pixels) and thus can properly cope with the above-mentioned situation,
at the expense of having to do $N_{\rm pix}$ neighbor searches per SN (which is never the bottleneck in simulations).

\citet{2018MNRAS.477.1578H} demonstrated that the properties of their simulated galaxies can differ dramatically by switching from the simple inject scheme (that suffers from the anisotropic problem) to their improved scheme.
They reported that the difference is more pronounced at high resolution ($m_{\rm gas} = 7\times 10^3 {\rm M_\odot}$) than at low resolution ($m_{\rm gas} = 5.6\times 10^4 {\rm M_\odot}$) as the anisotropic error in velocity increases for the same momentum budget when the injection region becomes smaller.
In contrast,
our improved scheme makes little difference to the galactic winds, especially at high resolution. 
This is not necessarily a conflict with \citet{2018MNRAS.477.1578H} as we are exploring a much higher resolution range.
In our high-resolution run ($m_{\rm gas} = 5{\rm M_\odot}$),
all SNe are properly resolved and 72\% of energy is injected as thermal energy which does not suffer from the anisotropic problem.
Although 28\% of energy is still injected as kinetic energy,
the difference between the two schemes turns out to be negligible.
In our lowest-resolution run ($m_{\rm gas} = 125{\rm M_\odot}$), 
the difference becomes the largest,
but in this case both schemes result in almost no winds anyway,
and thus the anisotropic problem is hardly a concern (compared to the lack of hot gas).

\subsection{Injecting terminal momentum as a sub-grid model for SN feedback?}

In large-scale cosmological simulations,
SN feedback will be necessarily unresolved and one always needs to adopt a sub-grid model.
Our numerical experiments 
suggest that while injecting $p_{\rm term}$ as a sub-grid model can be a viable way to drive turbulence in the ISM and regulate star formation \citep{2013ApJ...776....1K, 2015MNRAS.450..504M, 2015MNRAS.451.2900K},
it is not a suitable sub-grid model when it comes to launching galactic winds,
as its underlying assumption is that thermal energy has already been radiated away right after the injection.
Such an assumption leads to a huge underproduction of hot gas,
which has been shown to be the dominant population that can be launched as winds in many SNe-resolved stratified-box simulations \citep{2016MNRAS.456.3432G, 2017MNRAS.466.1903G, 2017ApJ...841..101L, 2017MNRAS.470L..39F, 2018ApJ...853..173K} as well as in  our global-disk simulations. 
In other words,
this sub-grid model breaks down because SN winds are pressure-driven rather than momentum-driven.

In principle, it is possible to generate some hot gas with $p_{\rm term}$-injection if the cumulative terminal momentum from a group of SNe becomes strong enough to shock-heat the ISM.
In practice, however,
this mechanism seems to be very inefficient in dwarf galaxies like ours.
In fact,
in our simulations, $p_{\rm term}$-injection generates even less hot gas (and hence weaker winds) than pure thermal injection.
This should not be too surprising.
After all,
this sub-grid model is designed to recover the correct terminal momentum rather than generating the right amount of hot gas.
On the other hand,
in a solar-neighborhood setup,
\citet{2017ApJ...846..133K} showed that the amount of hot gas in the ISM actually increases (and so does the outflow rate) as they degrade the resolution to cell size $\Delta x > 16$pc and stop seeing convergence (their Fig. 17).
This overproduced hot gas is not due to shock-heating via $p_{\rm term}$-injection.
Instead, it is due to their sink particle approach which suffers from artificially enhanced SN clustering at low resolution.
This leads to a blown-out of the entire disk,
and the subsequent SNe then become ``resolved'' (i.e. thermal energy is injected)
which give rise to the hot gas-dominated ISM due to the so-called ``thermal runaway'' \citep{2015MNRAS.449.1057G}.


We stress that we are not advocating using pure thermal injection as a sub-grid model to launch winds,
as it suffers from the well-known over-cooling problem.
Our results simply suggest that both schemes fail to launch winds if SNe are unresolved.
It remains unclear how to devise a sub-grid model that can not only inject the right amount of momentum into the ISM but also properly generate hot gas that can be launched as winds.


\subsection{Resolution requirement for convergence}\label{sec:resreq}

In our simulations, the wind properties converge at $m_{\rm gas} = 5 {\rm M_\odot}$.
This exact number may depend on the number of neighboring particles affected by an SN ($N_{\rm inj}$).
We adopt $N_{\rm inj} = 96$ in order to ensure that the injected region is properly resolved by at least one resolution element.
In SPH, this corresponds to the particle number in a smoothing kernel $N_{\rm ngb}$ (=100 in this work).
Such a large $N_{\rm ngb}$ (and hence $N_{\rm inj}$) is required in order to suppress the numerical noise and improve convergence in SPH \citep{2012MNRAS.425.1068D}.
More aggressive choice (smaller) of $N_{\rm inj}$ can be adopted,
though at the expense of not properly resolving the injected region,
which may trigger numerical fluid instabilities.
We explore the choice of $N_{\rm inj}$ and its effect on wind properties in Appendix \ref{app:Ninj}.
In other particle codes such as the meshless-finite-mass method \citep{2015MNRAS.450...53H},
it is reasonable to use a lower $N_{\rm inj}$ as the method has been shown to perform well with fewer neighbors (the standard value is $N_{\rm ngb} = 32$).
Another subtlety is whether energy/momentum is distributed to neighboring particles uniformly or weighted with the kernel function.
If a kernel weighting is adopted,
the effective $N_{\rm ngb}$ would be even smaller as most energy/momentum will be distributed to the nearest neighbors that have higher weightings.
This means that the resolution requirement (Eq. \ref{eq:coolmass}) can actually be less stringent.
We chose to be conservative by using a large $N_{\rm inj}$ in this work,
and thus our convergence requirement ($m_{\rm gas} = 5 {\rm M_\odot}$) can be viewed as a lower limit.
However,
our claim that one must resolve individual SNe to generate hot gas and launch winds should remain qualitatively robust.

In practice, it is not enough to only marginally satisfy the resolution requirement in Eq. \ref{eq:coolmass}.
In our 25M$_\odot$-run,
despite resolving 70\% of SNe,
the SN environments failed to converge (see Fig. \ref{fig:snrhot}).
This is because a marginally resolved SN defined as Eq. \ref{eq:coolmass} is still unable to resolve the diffuse bubble and dense shell of an SNR,
leading to a smoothed-out density structure.
When subsequent SNe occur within this no-so-diffuse bubble,
they will not be as efficient as they physically should be,
which in turn weakens the winds.
Our results suggest that the cooling mass needs to be resolved by around 5 injection mass,
which is consistent with other grid-based simulations \citep{2015ApJ...809...69S, 2015ApJ...802...99K} who show that the cooling radius needs to be resolved by at least a few cells.
\textbf{
We expect a similar resolution requirement (modulo the slightly larger $N_{\rm inj}$ required by SPH) would apply for other Lagrangian methods such as MFM and the moving mesh codes, though this still needs to be explicitly tested.
Indeed,
\citet{2018MNRAS.478..302S} who use the moving mesh code {\sc Arepo} \citep{2010MNRAS.401..791S} show that in their galactic scale simulations even with $m_{\rm gas} = 20 {\rm M_\odot}$ their wind properties are still not converged.
}


\textbf{
Intriguingly,
the recent FIRE-2 cosmological ``zoom-in'' simulations presented in \citet{2018MNRAS.477.1578H},
show that their stellar mass evolution of similar dwarf galaxies is well converged at $m_{\rm gas} = 2000 {\rm M_\odot}$,
which appears to be in conflicted with the results in both \citet{2018MNRAS.478..302S} and in this work.
}
The reason for the discrepancy requires further investigation, but a few points are worth noting:
(i) The convergence of stellar mass does not necessarily imply the convergence of wind properties, though these two are expected to be closely related.
(ii) At $m_{\rm gas} = 2000 {\rm M_\odot}$, most SNe in the FIRE-2 simulations are far from resolved (even with $N_{\rm inj} = 1$).
Therefore, 
the fact that the convergence can still be achieved suggests
that the galactic winds in the FIRE-2 simulations may not be driven by SNe but by other included feedback processes such as photoionization, stellar winds and radiative pressure\footnote{We note that these feedback processes are all treated by simplified recipes (the same applies to the photoionization treatment in our model) and further investigations are required to understand their effects.}.

\section{Summary}\label{sec:summary}

We have conducted high-resolution smoothed-particle hydrodynamics simulations ($m_{\rm gas} =  1{\rm M_\odot}$) of an isolated dwarf galaxy with a self-consistent ISM model which includes self-gravity, non-equilibrium cooling and heating, an H$_2$ chemistry network, individual star formation, stellar feedback and metal enrichment.
We study the properties of SN-driven winds of the galaxy and investigate the convergence of our model.
Our main results are summarized as follows.

\begin{enumerate}	
	
	\item Our self-consistent ISM model naturally leads to clustered SNe.
	Around 60\% of SNe occur within the preexisting hot bubbles (Fig. \ref{fig:snrhot}), which greatly enhances their dynamical impact on the ISM and leads to the formation of superbubbles with sizes of a few hundred pc (Fig. \ref{fig:ismfaceedgestars0399}).
	These superbubbles are able to break out of the disk, venting out hot and over-pressurized gas,
	which becomes the driving force of winds.	
	
	\item	
	In steady state,
	the gas flows can be characterized by the fountain flows and the winds.
	The fountain flows dominate (in mass) at small $r_s$ but they rapidly decline with $r_s$.
	At $r_s > 2$kpc,
	the fountain flows vanish and the actual winds emerge with nearly constant loading factors all the way to $R_{\rm vir}$.
	We find $\eta_m^{out} \sim 3$, $\eta_p^{out} \sim 1$ and $\eta_e^{out} \sim 0.05$ at $R_{\rm vir}$ (Fig. \ref{fig:mpeloadingz}).
	As a significant amount of gas in the ISM is blown out of the disk,
	the winds are only slightly metal-enriched ($y_Z^{out} \sim 1.5$ at $R_{\rm vir}$) due to the dilution by the unenriched ISM.
	The corresponding metal loading factor is $\eta_Z^{out} \sim 0.2$.
	The predicted winds are much weaker than what is typically adopted in cosmological simulations,
	which implies either our model is missing important feedback mechanisms or the winds in cosmological simulations need to be revised down.

	\item
	At $r_s < 2$kpc,
	most mass and momentum are carried by the warm fountain flows,
	while most energy is carried by the outflowing hot gas (Fig. \ref{fig:mpeloadingzphases}).
	After the winds are launched ($r_s > 2$kpc),
	they can be separated into two populations depending on their launching temperatures (Fig. \ref{fig:historywarm} and \ref{fig:historyhot}).	
	The hot gas acquires very high velocity (larger than $v_{\rm esc}$) after the breakout of SN bubbles and is able to escape the halo on its own,
	with a nearly constant Bernoulli parameter (or slightly decreasing as it feeds energy to the warm gas).
	On the other hand, the warm gas acquires velocity lower than $v_{\rm esc}$ at launching, but is later on accelerated and shock-heated by the subsequently launched hot winds, and eventually obtains enough energy to escape the halo.
	Therefore,
	its Bernoulli parameter keeps increasing with $r_s$.
	This highlights the caveat of extrapolating properties of warm gas from the launching radius to $R_{\rm vir}$ in small-box simulations.

	\item The wind properties converge at $m_{\rm gas} = 5 {\rm M_\odot}$ where the injection mass is $500 {\rm M_\odot}$ (Fig. \ref{fig:mloadconv}).
	This corresponds to the resolution where the cooling mass (Eq. \ref{eq:coolmass}) of individual SNe can be resolved by at least 5 resolution elements, or 5 kernel masses (Fig. \ref{fig:snrhot}).
	When SNe are unresolved, no hot gas can be generated and therefore no winds will be launched (Fig. \ref{fig:pd2by2converge}).

	\item 
	When SNe are properly resolved, winds are insensitive to the injection scheme of SN feedback (Fig. \ref{fig:mloadinject}).
	Injecting 100\% thermal energy leads to the same wind properties as injecting a mixture of thermal and kinetic energy based on the exact Sedov-Taylor solution.
	In addition, our new pixel-by-pixel momentum injection scheme with improved isotropy based on the {\sc HealPix} method has little effect on winds when SNe are resolved.
	
	\item 
	When SNe are unresolved, injecting thermal energy suffers from the overcooling problem as expected. 
	Injecting the terminal momentum $p_{\rm term}$, a popular sub-grid model in the literature, also fails to capture winds (injecting the residual thermal energy does not help, as shown in Appendix \ref{app:resEth}).
	This is because winds are driven by thermal pressure carried by the hot gas during the breakout of SN bubbles,
	rather than driven by the momentum carried mostly by the warm gas. 
	Since $p_{\rm term}$-injection assumes that thermal energy has already been radiated away, 
	it is unable to generate hot gas to launch winds (Fig. \ref{fig:coldbubble}).
	Further investigations are required to devise a sub-grid model that can not only generate turbulence in the ISM and regulate star formation but also produce enough hot gas to launch winds.

\end{enumerate}

\section*{Acknowledgments}
I thank the referee for the useful comments on the paper.
I would also like to thank Greg Bryan, Chang-Goo Kim, Miao Li, Mordecai Mac Low, Thorsten Naab and Phil Hopkins for enlightening discussions,
and Bernhard R\"ottgers for his help on {\sc pygad}\footnote{https://bitbucket.org/broett/pygad}, which I use for visualization in this work.
The Center for Computational Astrophysics is supported by the Simons Foundation.

\bibliographystyle{mn2e}
\bibliography{literatur}

\appendix
\section{Effect of the residual thermal energy} \label{app:resEth}
\begin{figure}
	\centering
	\includegraphics[width=0.99\linewidth]{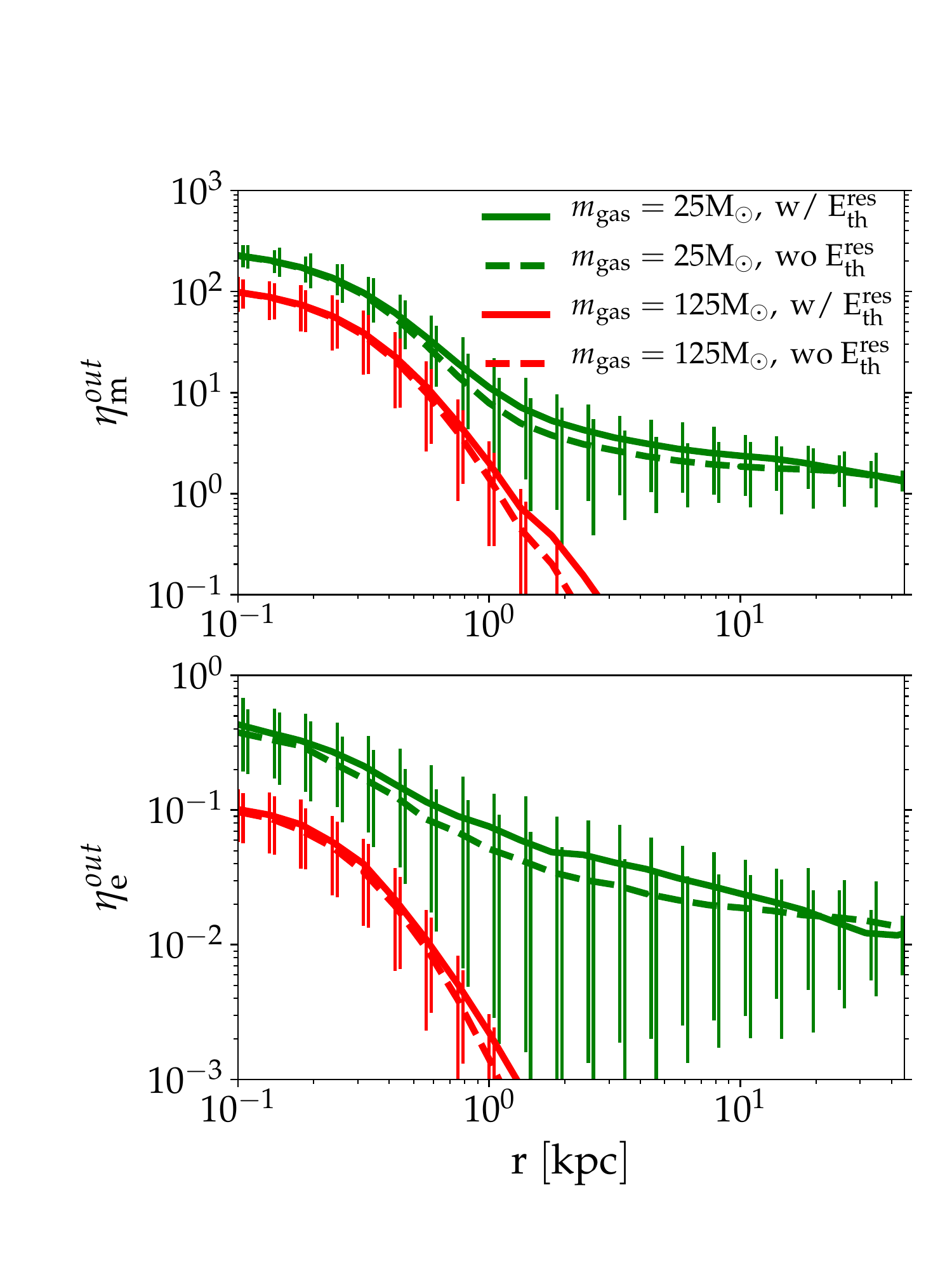}
	\caption{$\eta^{out}_m$ (top panel) and $\eta^{out}_e$ (bottom panel) as a function of $r_s$ for the 25${\rm M_\odot}$- and 125${\rm M_\odot}$-runs, with (solid lines) and without (dashed lines) injecting the residual thermal energy.
	The vertical error bars indicate the temporal fluctuations between 16 and 84 percentiles. Wind loading factors are insensitive to whether or not the residual thermal energy is included.}
	\label{fig:comparereseth}
\end{figure}

As discussed in Section \ref{sec:subgridSN},
injecting the terminal momentum of the SNR has become a widely-used sub-grid model in the literature.
However,
the residual thermal energy is not always included (see, e.g., \citealp{2015MNRAS.449.1057G, 2015MNRAS.454..238W, 2017ApJ...846..133K}).
In this appendix, 
we investigate the effect of the residual thermal energy by repeating the 25${\rm M_\odot}$- and 125${\rm M_\odot}$-runs without injecting $E^{\rm res}_{\rm th}$.
In Fig. \ref{fig:comparereseth}, we show $\eta^{out}_m$ (top panel) and $\eta^{out}_e$ (bottom panel) as a function of $r_s$ for these two runs (dashed lines) along with the runs that include $E^{\rm res}_{\rm th}$ (solid lines).
The vertical error bars indicate the temporal fluctuations between 16 and 84 percentiles.

The residual thermal energy has a negligible effect on the wind loading factors.
This is not surprising as thermal energy drops rapidly after cooling kicks in.
In addition,
since the unresolved SN is not able to clear out the gas and form a low-density bubble,
any residual thermal energy is forced to be injected in a dense region instead of in the low-density bubble as it physically should be,
and will therefore be radiated away even more rapidly for numerical reasons.
As a result,
it has little impact on the wind properties.

\section{Effect of the injection mass} \label{app:Ninj}
\begin{figure}
	\centering
	\includegraphics[width=0.99\linewidth]{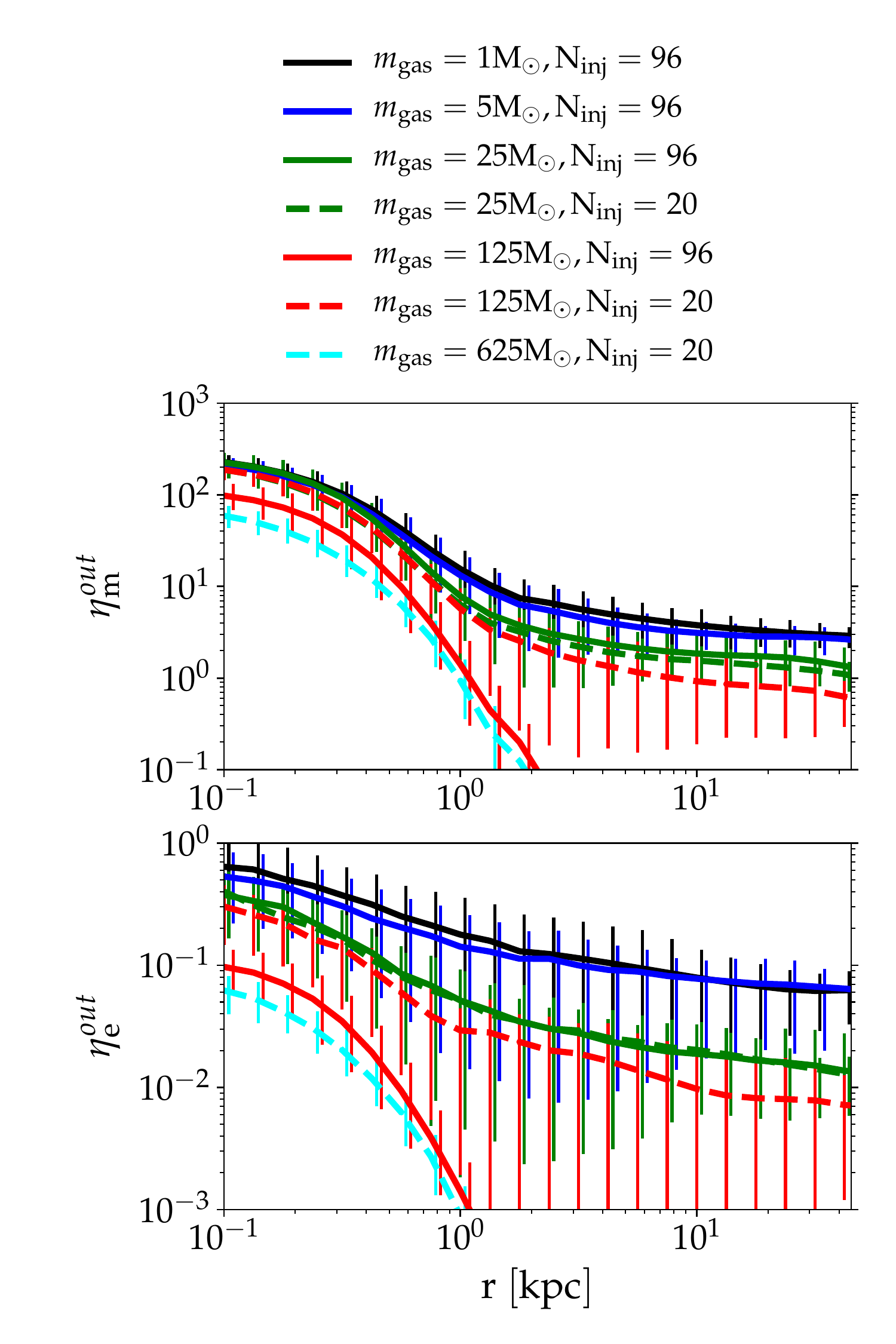}
	\caption{Same as Fig. \ref{fig:mloadconv} but overplotting with three new runs with 5 times smaller injection mass (dashed lines) and $m_{\rm gas} = 25, 125$ and $625 {\rm M_\odot}$, respectively.
	Lowering $N_{\rm inj}$ does increase the transition particle mass where winds weaken dramatically (from $m_{\rm gas} = 125{\rm M_\odot}$ to $625 {\rm M_\odot}$),
	which is defined by Eq. \ref{eq:coolmass}.
	However, the particle mass required for converged results ($m_{\rm gas} = 5{\rm M_\odot}$) does not increase with a smaller $N_{\rm inj}$.
	}
	\label{fig:meloadingninj}
\end{figure}

As discussed in Section \ref{sec:resreq},
we adopt a conservative choice of $N_{\rm inj} = 96$ to make sure that the injection region comparable to the SPH resolution and hence is properly resolved.
In this appendix,
we explore a more aggressive (around 5 times smaller) choice of $N_{\rm ing} = 20$ for the non-converged runs and test whether the convergence can be improved.
In Fig. \ref{fig:meloadingninj},
we show the same plot as Fig. \ref{fig:mloadconv} but overplotting with three new runs with 5 times smaller injection mass (dashed lines) and $m_{\rm gas} = 25, 125$ and $625 {\rm M_\odot}$, respectively.

The 25${\rm M_\odot}$-run is insensitive to the choice of $N_{\rm inj}$ and underpredicts $\eta^{out}_m$ and $\eta^{out}_e$ by a factor of 3 and 5 at $R_{\rm vir}$, respectively.
Note that the 25${\rm M_\odot}$-run with $N_{\rm inj} = 20$ has nearly the  same $M_{\rm inj}$ as the 5${\rm M_\odot}$-run with $N_{\rm inj} = 96$,
but the former is not converged while the latter is.
This is because, despite having the same $M_{\rm inj}$,
the one that has poorer SPH resolution will not be able to resolve the gas structure (e.g. diffuse bubbles, dense shells and low density channels).
As such,
it is dangerous to presume that convergence can be achieved at $x$-times higher $m_{\rm gas}$ just by simply adopting a $x$-times smaller $N_{\rm inj}$.

The 125${\rm M_\odot}$-run shows a significant stronger winds when a smaller $N_{\rm inj}$ is adopted,
though both $\eta^{out}_m$ and $\eta^{out}_e$ are still a factor of 5 and 10 lower than the converged values, respectively.
This is because with this injection mass most of the SNe are marginally resolved (cf. Fig. \ref{fig:snrhot}).
Meanwhile, the 625${\rm M_\odot}$-run has almost no winds.
This is in line with our conclusion that winds weaken dramatically once the SNe become unresolved,
and by adopting a smaller $N_{\rm inj}$, 
the transition scale is shifted towards higher mass scale.

To sum up,
lowering $N_{\rm inj}$ does increase the transition particle mass where winds weaken dramatically (from $m_{\rm gas} = 125{\rm M_\odot}$ to $625 {\rm M_\odot}$),
which is defined by Eq. \ref{eq:coolmass}.
However,
the particle mass required for converged results ($m_{\rm gas} = 5{\rm M_\odot}$) does not increase with a smaller $N_{\rm inj}$.

\end{document}